\documentclass[a4paper,fleqn]{cas-dc}

\usepackage[numbers,sort&compress]{natbib}

\usepackage{caption}
\usepackage{pifont} 
\usepackage{threeparttable}

\usepackage{hyperref}
\hypersetup{colorlinks=true, citecolor=blue, linkcolor=blue, urlcolor=blue}

\usepackage{wrapfig} 
\usepackage{flushend}
\usepackage{appendix}
\usepackage{bbding}
\usepackage{stmaryrd}

\def\tsc#1{\csdef{#1}{\textsc{\lowercase{#1}}\xspace}}
\tsc{WGM}\label{key}
\tsc{QE}
\tsc{EP}
\tsc{PMS}
\tsc{BEC}
\tsc{DE}

\begin{document}
\let\WriteBookmarks\relax
\def\floatpagepagefraction{1}
\def\textpagefraction{.001}

\shorttitle{PHA-Net: Prototype-based Hierarchical Alignment  Network for Text-Video Retrieval}

\shortauthors{Xiaolun Jing et al.}

\title [mode = title]{PHA-Net: Prototype-based Hierarchical Alignment  Network for Text-Video Retrieval}
	
\author[1,2]{Xiaolun Jing}[orcid=0000-0002-1159-062X]

\ead{jingxiaolun@sjtu.edu.cn}

\credit{Conceptualization of this study, Methodology, Experiment}

\address[1]{Ningbo Artificial Intelligence Institute, Shanghai Jiao Tong University, China}
\address[2]{School of Automation and Intelligent Sensing, Shanghai Jiao Tong University, China}

\author[1,2]{Kezhao Yin}[orcid=0009-0009-5017-204X]

\ead{ink2022@sjtu.edu.cn}

\address[3]{Department of Biomedical Engineering, Oregon Health \& Science University, Portland, OR, USA}
\author[3]{Xinxing Yang}[orcid=0000-0002-1512-2970]
\cormark[1]

\ead{yangxin@ohsu.edu}

\author[1,2]{Genke Yang}[orcid=0000-0003-3492-0211]
\cormark[1]
\cortext[1]{Corresponding author}


\ead{gkyang@sjtu.edu.cn}

\author[1,2]{Jian Chu}[orcid=0000-0002-8311-3419]

\ead{chujian@sjtu.edu.cn}

\begin{abstract}
With the emergence of large-scale image-text pre-training models, \textit{e.g.,} CLIP, text-video retrieval has experienced substantial advances in recent years. Existing best-performing methods involve aligning cross-modal semantics at individual, local, and global levels simultaneously, raising concerns about the intrinsic semantic mismatch between concise texts and rich videos. A canonical approach is to integrate multiple language-video attention modules into the hierarchical framework while this paradigm only optimizes visual representations with prohibitive computational costs. In this paper, we propose a new prototype-based hierarchical alignment network (PHA-Net) to align individual/local/global level representations across modalities. Concretely, we introduce multiple modality-shared prototypes as the bridge to efficiently optimize text and video representations for cross-modal alignment. Then, we argue that the imbalanced semantic distribution in clustered tokens may undermine retrieval performance, as tokens with weak semantics are of little interest. To reduce the impact of these tokens, a proposed prototype-supported token merge module is responsible for enhancing tokens with strong semantics and suppressing others with weak semantics via prototype semantics guidance. Moreover, we devise a prototype contrastive loss to encourage textual and visual prototypes to focus on different semantic information. The idea of this auxiliary loss is to ensure higher similarity between textual and visual prototypes from the same prototype than those from different prototypes. Extensive experiments on four benchmarks confirm the effectiveness of our PHA-Net, which achieves significant improvements in the sum of all recalls on MSR-VTT (8.8\%), ActivityNet (19.2\%), VATEX (0.7\%), and Charades (4.9\%). Code is available at \href{https://github.com/JingXiaolun/PHA-Net}{https://github.com/JingXiaolun/PHA-Net}.
\end{abstract}


\maketitle

\section{Introduction}
\label{Introduction} 
In recent years, the booming development of portable filming devices and video media platforms has led to a vast amount of video content. Searching videos of interest with the text query, referred to as text-video retrieval (TVR), has attracted increasing attention due to its high research and practical value. The recent progress in TVR is mostly driven by large-scale text-image pretrained model CLIP \cite{radford2021learning}, which has achieved strong performance and powerful generalization across numerous downstream cross-modal tasks like image captioning \cite{xu2015show}  and visual question answering (VQA) \cite{antol2015vqa}. As a direct extension of CLIP \cite{radford2021learning}, CLIP4Clip \cite{luo2022clip4clip} finetunes CLIP \cite{radford2021learning} and devises temporal fusion mechanisms to aggregate the features from different video frames for global alignment. Later, X-Pool \cite{gorti2022x} employs a parametric pooling scheme to generate global video feature with text-guided attention weights. However, simply pooling all frames as a whole expression may result in detailed visual information loss, affecting the match with specific textual entities. Therefore, existing methods adopt fine-grained TVR paradigms to conduct detailed semantic alignments. For instance, TokenFlow \cite{zou2022tokenflow} proposes a universal model-agnostic scheme for token-wise similarity calculation. DRL \cite{wang2022disentangled} explores pair-wise correlations between frames and words via the Weighted Token-wise Interaction (WTI). Although these methods have achieved impressive results, they are still far from simultaneously considering the contextual information (coarse-grained) and specific details (fine-grained) during text-video matching. Subsequently, X-CLIP \cite{ma2022x} innovatively proposes cross-grained contrasts (\textit{i.e.,} video-word and sentence-frame)
to filter out the unnecessary words and frames. UCoFiA \cite{wang2023unified} takes advantage of a multi-grained alignment strategy across patch-word, frame-sentence and video-sentence levels. Despite their success, they only rely on either individual-level (frames/words) or global-level (video/sentence) information to perform the cross-modal alignment without taking local-level (clips/phrases) information into account. Intuitively, both text and video contain three complementary components (words/phrases/sentence and frames/clips/video) ranging from individual level to global level, and the usage of individual-to-global semantic alignments is critical to achieve accurate retrieval.

As a pioneering work, HCMI \cite{jiang2022tencent} first introduces local-level alignment (phrase-clip) into TVR for hierarchical cross-modal interactions, bringing clear performance gains on both text-to-video and video-to-text retrieval tasks. HBI \cite{jin2023video} further constructs a hierarchical banzhaf interaction framework to value possible correspondence between text words and video frames. Though effective, the intrinsic semantic misalignment between individual level features (\textit{i.e.,} words and frames) is ignored in the above methods. Frames tend to contain a much wider gamut of information, with certain frames being entirely irrelevant to words, thereby posing challenges for hierarchical cross-modal alignment. To this end, a feasible solution is to produce semantic-relevant frame features through the transformer-based cross-attention module \cite{gorti2022x}. However, this module encounters two issues: \textbf{(1) Visual-only Optimization.} The cross-attention module purely uses text-guided attention weights to generate semantic-related video feature without optimizing the text counterpart. Thus, the trained model may fail to handle the text-video matching involving polysemous words or complex semantics. \textbf{(2) Prohibitive Computational Complexity.} At least one cross-attention module is required to narrow the modality gap and establish hierarchical  correspondence, leading to significantly slower inference speed. Thus, how to jointly optimize hierarchical text and video representations in a relatively efficient manner remains to be explored. 

To this end, we propose a \textbf{P}rototype-based \textbf{H}ierarchical \textbf{A}lignment \textbf{Net}work (PHA-Net) to align hierarchical representations across different modalities in a prototype-guided manner. Specifically, we first introduce a set of modality-shared prototypes at individual level and local level to capture the latent shared semantics across cross-modal features, effectively bridging the modality gap as well as achieving the joint optimization of textual and visual features. Notably, on the one hand, the learnable prototypes improve the training convergence speed. On the other hand, they incur acceptable computational overhead in the inference stage.
Next, building upon the semantic-aligned individual-level tokens, we further design a prototype-supported token merge module to handle the imbalanced semantics distribution among clustered tokens. Due to the limited number of prototype tokens, only some clustered tokens 
contain the shared prototype semantics while others lack prototype information, thereby leading to unfair semantic distribution. In detail, since tokens with strong semantics are critical and others are of little interest, the significance degree of different clustered tokens must be taken into account.
To accomplish this, we use prototype-guided attention weights over the clustered tokens to enhance tokens with strong semantics and suppress others with weak semantics. Note that our prototypes are averaged and repeated along the token dimension to match the clustered tokens' dimensions. By stacking prototype-supported merging modules, we can get the local-level (phrase/clip) and global-level (sentence/video) tokens for cross-modal alignment at higher levels.
Furthermore, to ensure semantic diversity among shared prototypes, we devise a prototype contrastive loss (PCL), which constrains the output textual and visual prototypes from the same prototype to be more similar than those from different prototypes. 
With the auxiliary prototype contrastive loss optimization, our model alleviates the semantics overlapping issue and meanwhile fosters a rich information exploration in text and video.

To summarize, our contributions are as follows: \par
$\bullet$ We propose a prototype-based hierarchical alignment text-video retrieval network to achieve cross-modal alignment in a relatively efficient manner.

$\bullet$ We present three key designs: the modality-shared prototype, the prototype-supported token merge module and the prototype contrastive loss to align cross-modal semantics, facilitate the usage of prototype semantics and guarantee the prototype diversity.

$\bullet$ The proposed model achieves competitive performance with the mainstream approaches across four benchmark datasets of MSR-VTT \cite{xu2016msr}, ActivityNet \cite{caba2015activitynet}, VATEX \cite{wang2019vatex} and Charades \cite{sigurdsson2016hollywood}.

The remainder of this paper is organized as below: Section \ref{Sec: Related Work} provides a review of related works. Section \ref{Sec: Methodology} elaborates on the details of PHA-Net. Section \ref{Sec: Experiments} presents the experimental setup and analyzes the experimental results, with ablation studies to validate the effectiveness of our approach. Section \ref{Sec: Conclusion} concludes the paper.

\section{Related Work}
\label{Sec: Related Work}

\subsection{Cross-Modal Retrieval}
With the explosive growth of diverse multimodal data, cross-modal retrieval has encompassed a broader array of  modalities and tasks, including text-image \cite{ma2023fedsh, zeng2026towards, ma2025tsgr2},  text-audio \cite{xin2023improving, oncescu2024sound}, text-video \cite{ luo2022clip4clip, meng2026spatio}, and others. For text-image, FedSH \cite{ma2023fedsh} builds a self-aligned network to tackle the local model generalization and entity boundary obscuring limitations. FNM \cite{zeng2026towards} presents a false negative mitigation framework to facilitate feature alignment and decrease deviated identification. $\text{TSGR}^{2}$ \cite{ma2025tsgr2} employs multi-level relation reasoning and adaptive language masking strategies for enhanced fine-grained alignment.  For text-audio, TAP \cite{xin2023improving} is a text-aware attention pooling module for more accurate retrieval. The authors of \cite{oncescu2024sound}  use Large Language Models (LLMs) to generate audio-centric descriptions. For text-video, CLIP4Clip \cite{luo2022clip4clip} devises three temporal aggregation schemes for coarse-grained matching. HSP-SA \cite{meng2026spatio} integrates structural priors into cross-modal learning to improve the reliability of person retrieval. In line with CLIP4Clip \cite{luo2022clip4clip}, our work is dedicated to universal text-video retrieval.  

\subsection{Text-Video Retrieval}
Text-video retrieval requires the model to establish the correct match between texts and videos. Canonical paradigm tends to obtain offline features via pre-trained text and video extractors. CE \cite{liu2019use}, MMT \cite{gabeur2020multi}, and HiT \cite{liu2021hit} are all such works. However, these methods employ additional experts to pre-process data, which increases the complexity and limits flexibility. Later, some works benefit from end-to-end solutions. ClipBERT \cite{lei2021less} and Frozen \cite{bain2021frozen} develops efficient end-to-end pre-training schemes. MCQ \cite{ge2022bridging} predicts verb or noun features to build cross-modal associations. Additionally, MCKD \cite{ma2023using} proposes a multimodal contrastive knowledge distillation method to calibrate mixed
boundaries.

Recently, due to the exceptional performance and powerful generalization of large-scale image-text pre-trained models, CLIP4Clip \cite{luo2022clip4clip} applies CLIP \cite{radford2021learning} as the backbone in text-video retrieval, achieving satisfying results and inspiring a series of works. X-Pool \cite{gorti2022x} attempts to aggregate frame features based on text-guided attention. X-CLIP \cite{ma2022x} proposes cross-grained contrasts to lower the impact of unnecessary clues. CenterCLIP \cite{zhao2022centerclip} and TS2-Net \cite{liu2022ts2} devise clustering and selection mechanisms to find the most representative tokens. EMCL-Net \cite{jin2022expectation} uses the expectation-maximization algorithm to find a compact set of bases for the latent space. DRL \cite{wang2022disentangled} explores fine-grained pair-wise correlations by a weighted token-wise interaction module. UATVR \cite{fang2023uatvr} models each text-video lookup as a distribution matching procedure. UCoFiA \cite{wang2023unified} accomplishes multi-grained alignment for coarse- and fine-grained matching. DiffusionRet \cite{jin2023diffusionret} develops a generative diffusion-based framework for text-video retrieval. EERCF \cite{tian2024towards} and TeachCLIP \cite{tian2024holistic} ensure effectiveness and efficiency through lightweight blocks and knowledge distillation. TC-MGC \cite{jing2025tc} introduces word-frame attention module to generate semantic-related frame representations for better multi-grained contrastive learning.

Albeit the promising advances, the above methods are limited to either individual-level or global-level alignments, while overlooking the local-level correspondence. HGR \cite{chen2020fine} builds a semantic graph to explore hierarchical relations. HANet \cite{wu2021hanet} disentangles videos into entity, action, and event levels for hierarchical alignment. QAMF \cite{ma2022query} develops a query-adaptive fusion mechanism to fuse multi-level semantic representations. Additionally, HBI \cite{jin2023video} innovatively designs hierarchical banzhaf interactions to explore cross-modal relations among entities, actions, and events. HCMI \cite{jiang2022tencent} pioneers the construction of multi-level feature representations and hierarchical cross-modal interactions. Nevertheless, these methods still suffer from the semantics mismatch between concise words and rich frames features, leading to a dilemma in achieving better performance. As a result, we aim to align hierarchical cross-modal representations via trainable modality-shared prototypes for bridging the inherent semantic gap.

\subsection{Prototype Learning}
Prototype learning is a popular learning approach that is usually employed to learn a set of representative prototypes for each category. Previous studies focus on exploiting prototype learning for uni-modal tasks, \textit{e.g.,} semantic segmentation \cite{wang2019panet}, object detection \cite{yan2019meta}, and object tracking \cite{wang2012online}. Recently, the integration of prototype learning into cross-modal tasks has achieved remarkable success for significant performance. Specifically, PLGA \cite{meng2022prototype} jointly performs the fine-grained local alignment and high-level global alignment in a prototype-based alignment network for image–text retrieval. PTSN \cite{zeng2022progressive} introduces tree-structured prototypes to model
the hierarchical semantic information of concepts on the image captioning model. Meanwhile, some recent works attempt to optimize text-video retrieval performance via prototype learning. For example, TMVM \cite{lin2022text} proposes to automatically generate multiple prototypes for video features aggregation. ProST \cite{li2023progressive} decomposes the matching process into complementary object-phrase and event-sentence prototype alignments for progressive spatio-temporal prototype matching. S2CA \cite{li2025s2ca} uses shared prototype mechanism to establish concept-level alignment between texts and videos. Different from them, we train our hierarchical retrieval model using several modality-shared prototypes for cross-modal alignment at individual level, local level, and global level, respectively.

\section{Methodology}
\label{Sec: Methodology}
\begin{figure*}[!t]
	\centerline{\includegraphics[width=\textwidth]{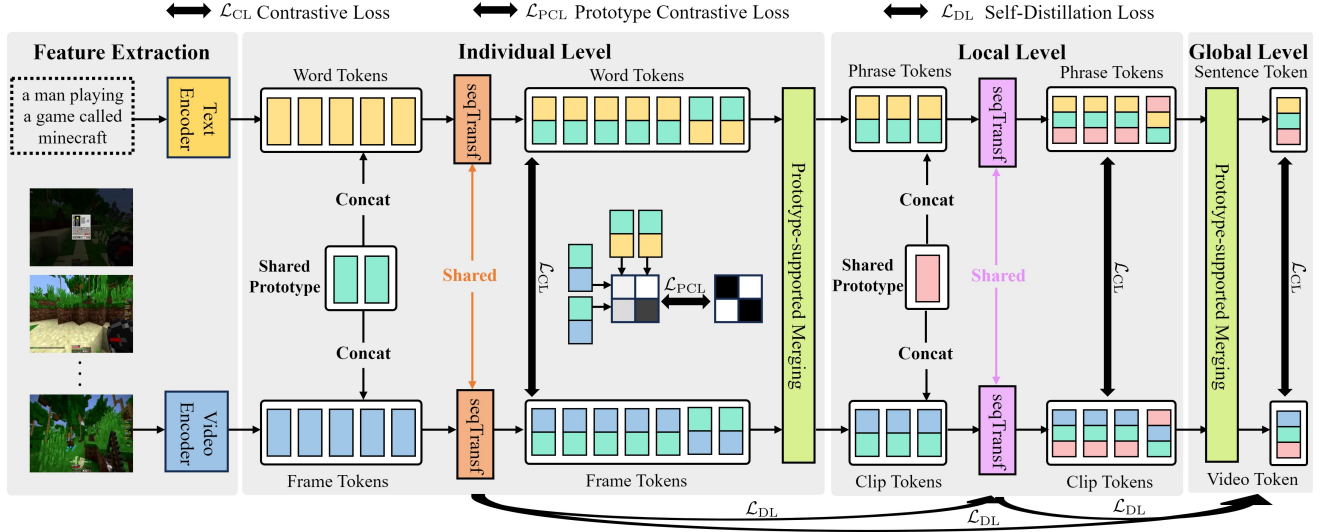}}
	\caption{\textbf{Overview of the proposed PHA-Net. } There are three main components in PHA-Net: (1) A set of modality-shared learnable prototypes for bridging the modality gap and facilitating hierarchical cross-modal alignments. (2) The prototype-supported token merge module for integrating prototype semantic guidance into the merging process. (3) The prototype contrastive loss for encouraging semantic diversity between textual and visual prototypes. Besides, we use the additional self-distillation loss among individual level, local level, and global level to improve the generalization ability.}
	\label{fig: main_architecture}
\end{figure*}

In this section, we present our Prototype-based Hierarchical Alignment Network (PHA-Net) in Fig. \ref{fig: main_architecture} for text-video retrieval. First, we outline the foundational concepts including feature extraction and fine-grained interactions in the Section \ref{Sec: Preliminaries}. Then, we describe how to achieve hierarchical alignment at individual level, local level and global level in the Section \ref{Sec: Hierarchical Alignment}. Next, we introduce our auxiliary prototype contrastive loss in the Section \ref{Sec: Prototype Contrastive Loss}.
Finally, we explain the training details and inference method in the Section \ref{Sec: Training and Inference}. 

\subsection{Preliminaries}
\label{Sec: Preliminaries}
\textbf{Problem Formulation.} We denote the text collection as $\mathcal{T}$ and the video collection as  $\mathcal{V}$. The objective of TVR is to learn a similarity function $s(\cdot)$, which maximizes the similarity score of relevant samples while minimizes similarity score of irrelevant pairs. Formally, given a text query $t \in \mathcal{T}$, we leverage the CLIP text encoder $\phi_{t}(\cdot)$ to output word features $T_{w}$, where $T_{w}=[w_{1},w_{2},\cdots,w_{N_{w}}] \in \mathbb{R}^{N_{w} \times D}$, $N_{w}$ is the length of words, and $D$ is the feature dimension. Similar to the text features extraction, we use the CLIP video encoder $\phi_{v}(\cdot)$ to produce visual features. Given a video $v \in \mathcal{V}$, we first uniformly sample $N_{f}$ video frames $v=[f_{1}^{o},f_{2}^{o},\cdots,f_{N_{f}}^{o}]  $. For the $k$-th frame of the video $f_{k}^{o}$, we divide it into fix-sized disjoint patches and prepend a [CLS] token to them. After that, the pre-processed video frames are fed into the video encoder 
$\phi_{v}(\cdot)$ to obtain the patch representation $p_{k}$, where $p_{k}=\phi_{v}(f_{k}^{o}) \in \mathbb{R}^{P \times D}$, $P$ indicates the number of patches within a video frame. We then concatenate the [CLS] representation from each frame together and get the frame features $V_{f}=[f_{1},f_{2},\cdots,f_{N_{f}}] \in \mathbb{R}^{N_{f} \times D}$.

The core of text-to-video retrieval is to rank all videos $v \in \mathcal{V}$ based on their semantic similarity $S_{t,v}$ to a given query text $t$, and vice versa for video-to-text retrieval. The semantic similarity is typically measured by the cosine similarity between the text and video features:
\begin{equation}
    S_{t,v}= \frac{t \cdot v}{\lVert t \rVert \cdot \lVert v \rVert}. \label{Eq. global similarity}
\end{equation}

In training, we adopt the symmetric InfoNCE loss to calculate the cross-modal contrastive loss, which can be formulated as:
\begin{equation}
	\begin{split}
    \mathcal{L}_{\text{CL}}=-\frac{1}{2}[\frac{1}{B} \sum_{i=1}^{B} \textup{log} \frac{\textup{exp}(S_{t_{i}, v_{i}}/\sigma)}{\sum_{j=1}^{B}\textup{exp}(S_{t_{i}, v_{j}}/\sigma)} + \\
   \frac{1}{B} \sum_{i=1}^{B} \textup{log} \frac{\textup{exp}(S_{t_{i}, v_{i}}/\sigma)}{\sum_{j=1}^{B}\textup{exp}(S_{t_{j}, v_{i}}/\sigma)}], \label{Eq: contrastive loss}
   \end{split}
\end{equation}
where $\sigma$ is the temperature factor, $B$ is the mini-batch size of text-video pairs, and $S_{t_{i}, v_{j}}$ refers to the similarity matrix between text $t_{i}$ and video $v_{j}$.

\textbf{Fine-grained Interactions.} Since the similarity between word and frame representations is a matrix, we use the widely-used weighted token-wise interaction mechanism to obtain text-to-video and video-to-text similarity, respectively. Following \cite{wang2022disentangled}, we get the maximum value of each row and each column, followed by the computed adaptive weights to obtain the similarity score in bi-directions. The instance-level similarity score thus comes from the average of above two similarities, formulated as:
\begin{equation}
    S_{t,v}=\frac{1}{2} \left(\underbrace{\sum_{m=1}^{N_{w}} \gamma_{t}^{m} \max\limits_{n=1}^{N_{f}} \langle w_{m}, f_{n} \rangle}_{\text{text-to-video similarity}} + \underbrace{\sum_{n=1}^{N_{f}} \gamma_{v}^{n} \max\limits_{m=1}^{N_{w}} \langle w_{m}, f_{n} \rangle}_{\text{video-to-text similarity}}\right), \label{Eq: fine_grained interaction}
\end{equation}
where $\langle \cdot, \cdot \rangle$ means the inner production function, $\gamma_{t}=\text{Softmax}(\text{MLP}_{t}(T_{w})) \in \mathbb{R}^{1 \times N_{w}}$ and $\gamma_{v}=\text{Softmax}(\text{MLP}_{v}(V_{f})) \in \mathbb{R}^{1 \times N_{f}}$ are the weights of text words and video frames, respectively. $w_{m}$ and $f_{n}$ are channel-wise normalized before computing similarity.

\subsection{Hierarchical Alignment}
\label{Sec: Hierarchical Alignment}

Generally, vanilla hierarchical alignment is initially established among individual level tokens extracted from respective encoders, and then extended to aggregated or clustered local level and global level tokens. However, hierarchical alignment in a simple encoder-dependent way does not consider the semantic mismatch problem. To tackle the problem, we introduce multiple modality-shared prototypes to assist cross-modal semantic alignment at three different levels, including individual level, local level and global level.

\textbf{Individual Alignment.} Given a set of word tokens $T_{w}=\{w_{m}\}_{m=1}^{N_{w}}$ and frame tokens $V_{f}=\{f_{n}\}_{n=1}^{N_{f}}$ from the output of backbone encoders, we first randomly initialize several modality-shared prototypes $P_{I}=\{p_{I}^{i}\}_{i=1}^{N_{I}^{p}}$ to encourage semantics sharing across words and frames. Then, we concatenate the extracted tokens and learnable prototypes together, followed by a modality-shared lightweight sequential transformer block (TB) to further model relations between prototypes and tokens. Note that the position embeddings are omitted for brevity. Taking the word side as an example, we can obtain the output word tokens $T_{w}^{'}=[w_{1}^{'}, w_{2}^{'},\cdots, w_{N_{w}+N_{I}^{p}}^{'}]$, $T_{w}^{'} \in \mathbb{R}^{(N_{w}+N_{I}^{p}) \times D}$ as follows:
\begin{equation}
	T_{w}^{'}=\text{TB}(\text{Concat}(w_{1},w_{2},\cdots,w_{N_{w}}, p_{I}^{1},p_{I}^{2},\cdots, p_{I}^{N_{I}^{p}})), \label{Eq: word_features}
\end{equation}
where Concat($\cdot$) is the concatenation operation. Similarly, the output frame tokens $V_{f}^{'} \in \mathbb{R}^{(N_{f}+N_{I}^{p}) \times D}$ can be acquired with the same formulation as Eq. \ref{Eq: word_features}. Next, we calculate the similarity $S_{t,v}^{I}$ by replacing $w_{m}$ ($f_{n}$) with $w_{m}^{'}$ ($f_{n}^{'}$) in Eq. \ref{Eq: fine_grained interaction}. Finally, the contrastive loss $\mathcal{L}_{\text{CL}}^{I}$ is calculated by replacing $S_{t,v}$ with $S_{t,v}^{I}$ in Eq. \ref{Eq: contrastive loss}.

\begin{figure}[!t]
	\centerline{\includegraphics[width=0.5\textwidth]{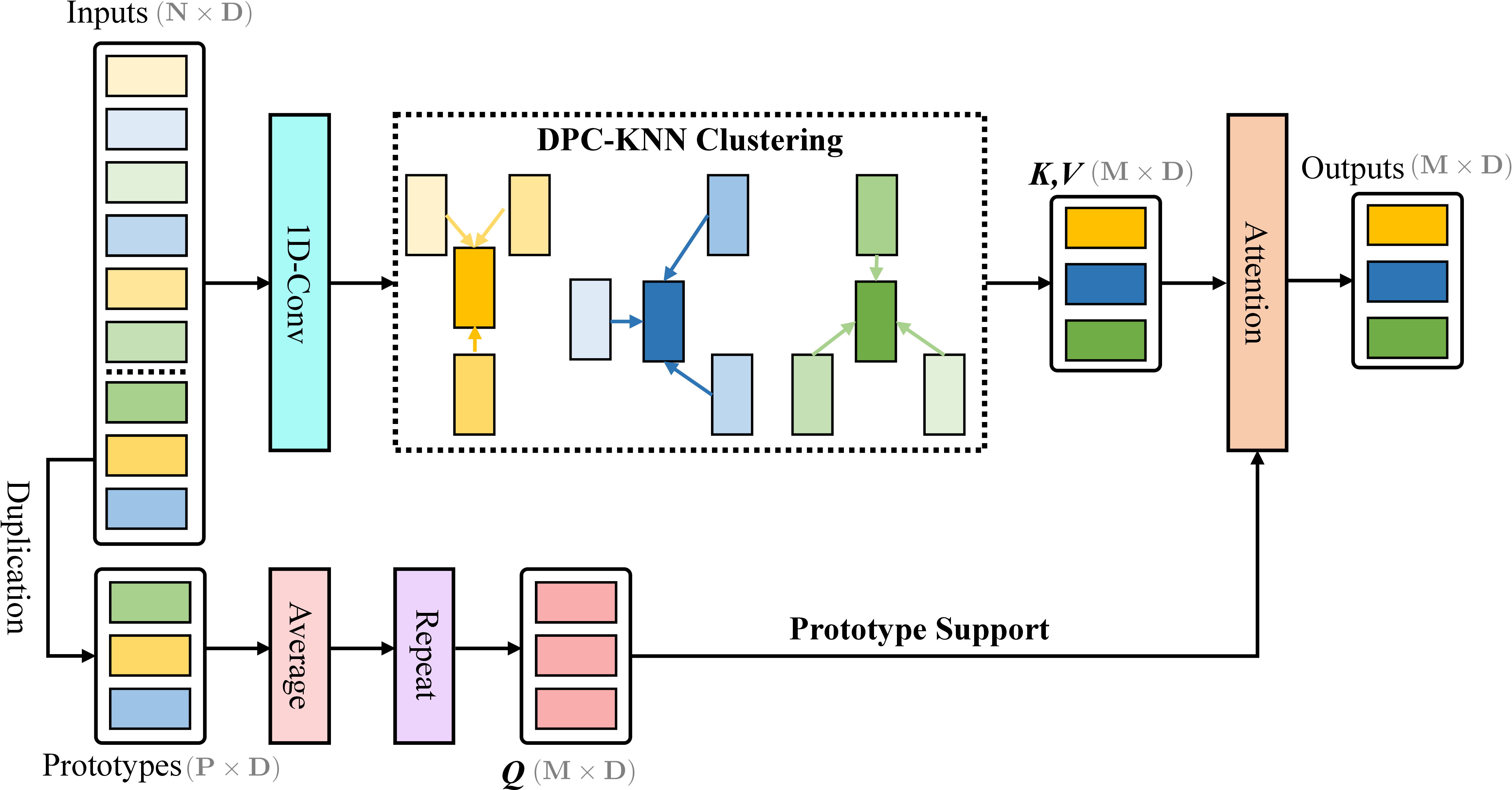}}
	\caption{\textbf{The prototype-supported token merge module.} ``1D-Conv'' means the one-dimensional convolutional layer. $N$ input tokens with $D$ channels are first clustered into $M$ clusters. Subsequently, $P$ prototype tokens in the inputs are averaged and repeated along the token dimension to get $M$ prototype tokens. Finally, we feed the obtained $M$ prototype tokens as queries $Q$ and the merged tokens as keys $K$ and values $V$ into an attention module. } 
	\label{fig: ptcb_cluster}
\end{figure}

\textbf{Local Alignment.} For local-level alignment, we follow \cite{jin2023video} to cluster the above word (frame) tokens into phrase (clip) tokens. However, since the number of prototypes is typically smaller than input words (frames), the direct clustering may assign shared prototype semantics to partial phrases (clips), thus resulting in unfair semantic distribution among the clustered phrase (clip) tokens. To some extent, phrase (clip) tokens with strong semantics deserve more attention.
Therefore, we devise a prototype-supported token merge module in Fig. \ref{fig: ptcb_cluster}, which leverages the prototype semantics guidance to enhance phrase (clip) tokens with strong semantics and suppress 
others with weak semantics.

Starting with the output word tokens $T_{w}^{'}$ in Eq. \ref{Eq: word_features}, we first utilize a one-dimensional convolutional layer to model the temporal relations among them. Then, we employ a k-nearest neighbor-based density peaks clustering algorithm, DPC-KNN \cite{du2016study}, to group the similar word tokens. Specifically, we compute the local density $\rho_{i}$ of each token based on its $K$-nearest neighbors:
\begin{equation}
	\rho_{i}=\text{exp}\left(-\frac{1}{K} \sum_{w_{k}^{'} \in \text{KNN}(w_{i}^{'})} \lVert w_{k}^{'}-w_{i}^{'} \rVert ^{2}\right),
\end{equation}
where $\text{KNN}(w_{i}^{'})$ is the $K$-nearest neighbors of token $w_{i}^{'}$. After that, we compute the distance index $\delta_{i}$ of each token $w_{i}^{'}$:
\begin{equation}
	\delta_{i}=
	\begin{cases}
		\underset{j:\rho_j>\rho_i}{\textrm{min}} \Vert w_{j}^{'}-w_{i}^{'} \Vert^2, & \text{if $\exists j$ s.t. $\rho_j>\rho_i$.}\\
		\ \ \underset{j}{\textrm{max}} \ \ \Vert w_{j}^{'}-w_{i}^{'} \Vert^2, & \text{otherwise.}
	\end{cases}
\end{equation}
Intuitively, higher $\rho$ denotes the token is in a high-density region and higher $\delta$ represents it is far from other high-density tokens. Therefore, we treat those tokens with higher $\rho_{i} \times \delta_{i}$ as cluster centers to merge other tokens into the nearest cluster center according to the Euclidean distances. Next, we regard the weighted average tokens of each cluster as the corresponding cluster, where the weight $W=\text{Softmax}(\text{MLP}_{w}(T_{w}^{'}))$. Finally, we take advantage of the prototype semantics to guide local-level phrase tokens generation. To achieve the same dimension as the weighted average tokens, the input prototypes are averaged and repeated along the token dimension. Meanwhile, we treat the generated prototype tokens as queries $Q$, and the weighted average tokens as keys $K$ and values $V$, with output from the cross-attention module being the prototype-supported phrase tokens :
\begin{equation}
	T_{p}=\text{Softmax}\left(\frac{QK^{T}}{\sqrt{D}}\right)V,
\end{equation}
where $T_{p} \in \mathbb{R}^{N_{p} \times D}$ are the phrase tokens, and $N_{p}$ denotes the number of phrases. Similarly, the output frame tokens $V_{f}^{'}$ are fed into the token merge module to obtain the clip tokens $V_{c} \in \mathbb{R}^{N_{c} \times D}$, where the DPC-KNN clustering is based on the semantic similarity and $N_{c}$ indicates the number of clips.

To better align the obtained phrase and clip semantics, we append $N_{L}^{p}$ modality-shared prototypes respectively at the end of phrase and clip tokens. Similar to above individual alignment, we use another modality-shared lightweight transformer block to enhance the temporal information between tokens and output the semantic-aligned phrase tokens $T_{p}^{'} \in \mathbb{R}^{(N_{p}+N_{L}^{p}) \times D}$ and clip tokens $V_{c}^{'} \in \mathbb{R}^{(N_{c}+N_{L}^{p}) \times D}$. Once the phrase and clip tokens are attained, we  substitute $w_{m}$ ($f_{n}$) with $p_{m}^{'}$ ($c_{n}^{'}$) in Eq. \ref{Eq: fine_grained interaction} to get the similarity $S_{t,v}^{L}$, which is used in Eq. \ref{Eq: contrastive loss} for contrastive loss $\mathcal{L}_{\text{CL}}^{L}$ calculation.

\textbf{Global Alignment.} In order to perform global alignment, we employ the prototype-supported token merge module to produce sentence token $T_{s}^{'} \in \mathbb{R}^{1 \times D}$ and video token $V_{v}^{'} \in \mathbb{R}^{1 \times D}$ from the obtained phrase and clip tokens, respectively. Then, we compute the similarity $S_{t,v}^{G}$ by replacing $t$ ($v$) with $T_{s}^{'}$ ($V_{v}^{'}$) in Eq. \ref{Eq. global similarity}, followed by the usage of $S_{t,v}^{G}$ in Eq. \ref{Eq: contrastive loss} to calculate the contrastive loss $\mathcal{L}_{\text{CL}}^{G}$.

\subsection{Prototype Contrastive Loss}
\label{Sec: Prototype Contrastive Loss}
Although we employ modality-shared prototypes for better cross-modal alignment, there may exist ``same semantics'' among generated textual and visual prototypes. To prevent the prototypes from focusing on the same region, we devise a prototype contrastive loss to maintain the semantic diversity within the textual and visual prototypes in the batch-level. Notably, this loss is only applied to multiple prototypes. Taking individual-level prototypes output from the transformer block as an example, with the word prototype features $P_{I}^{w} \in \mathbb{R}^{B \times N_{I}^{p} \times D}$ and frame prototype features $P_{I}^{f} \in \mathbb{R}^{B \times N_{I}^{p} \times D}$, we first reshape the prototype features to $\mathbb{R}^{N_{I}^{p} \times B \times D}$ and use the matrix multiplication to compute the similarity matrix $S \in \mathbb{R}^{N_{I}^{p} \times N_{I}^{p} \times B \times B}$. Then, we apply the fine-grained interaction operation in Eq. \ref{Eq: fine_grained interaction} along batch level to get a prototype similarity $S^{P} \in \mathbb{R}^{N_{I}^{p} \times N_{I}^{p}}$. Finally, we consider the InfoNCE loss to constrain the $i$-th word prototype close to its corresponding frame prototype and far from others:
\begin{equation}
	\mathcal{L}_{\text{PCL}}=-\frac{1}{N_{I}^{p}} \sum_{i=1}^{N_{I}^{p}} \textup{log} \frac{\textup{exp}(S_{i, i}^{P}/\sigma)}{\sum_{j=1}^{N_{I}^{p}}\textup{exp}(S_{i, j}^{P}/\sigma)}. \label{Eq: prototype contrastive loss}
\end{equation}

\subsection{Training and Inference}
\label{Sec: Training and Inference}
\textbf{Training.}
With the above cross-modal contrastive loss at three levels, we calculate the hierarchical contrastive loss as:
\begin{equation}
	\mathcal{L}_{\text{HCL}}=\mathcal{L}_{\text{CL}}^{I}+\alpha \mathcal{L}_{\text{CL}}^{L}+\beta \mathcal{L}_{\text{CL}}^{G}, \label{Eq: hierarchical contrastive loss}
\end{equation}
where $\alpha$ and $\beta$ are the weighting parameters, $\mathcal{L}_{\text{CL}}^{I}$, $\mathcal{L}_{\text{CL}}^{L}$, and $\mathcal{L}_{\text{CL}}^{G}$ represent the contrastive loss at individual level, local level, and global level, respectively.

In order to improve the generalization ability, we introduce self-distillation loss (DL) to optimize Kullback-Leibler (KL) divergence between the distribution among similarity matrix from different hierarchical levels, including
individual level to local level (I2L), local level to global level (L2G), and individual level to global level (I2G). For I2L distillation, we define the probability distribution $\mathcal{D}_{t,v}^{I}$ and $\mathcal{D}_{v,t}^{I}$ as follows:
\begin{align}
	\mathcal{D}_{t,v}^{I}&=[p_{i,1}, p_{i,2}, \cdots, p_{i,B}], i \in [1, B],\\
	\mathcal{D}_{v,t}^{I}&=[\hat{p}_{1,j}, \hat{p}_{2,j}, \cdots, \hat{p}_{B,j}], j \in [1, B],
\end{align}
where $p_{i,j}=\frac{\text{exp}(S_{t_{i}, v_{j}}^{I})}{\sum_{k=1}^{B} \text{exp}(S_{t_{i}, v_{k}}^{I})}$, 
$\hat{p}_{i,j}=\frac{\text{exp}(S_{t_{i}, v_{j}}^{I})}{\sum_{k=1}^{B} \text{exp}(S_{t_{k}, v_{j}}^{I})}$. Similarly, the probability distribution $\mathcal{D}_{t,v}^{L}$ and $\mathcal{D}_{v,t}^{L}$ are calculated in the same way using $S_{t,v}^{L}$. Thus, the $\mathcal{L}_{\text{DL}}^{I2L}$ loss is formulated as:
\begin{equation}
	\mathcal{L}_{\text{DL}}^{I2L}=\mathbb{E}_{t,v}[\text{KL}(\mathcal{D}_{t,v}^{L} \| \mathcal{D}_{t,v}^{I}) + \text{KL}(\mathcal{D}_{v,t}^{L} \| \mathcal{D}_{v,t}^{I})]. \label{Eq: I2L loss}
\end{equation}

\noindent The remaining $\mathcal{L}_{\text{DL}}^{L2G}$ loss and $\mathcal{L}_{\text{DL}}^{I2G}$ loss are calculated in the same way as Eq. \ref{Eq: I2L loss}. Therefore, the $\mathcal{L}_{\text{DL}}$ loss is derived from the sum of above three distillation losses:
\begin{equation}
	\mathcal{L}_{\text{DL}}=\mathcal{L}_{\text{DL}}^{I2L}+\mathcal{L}_{\text{DL}}^{L2G}+\mathcal{L}_{\text{DL}}^{I2G}. \label{Eq: distillation loss}
\end{equation}

By combining $\mathcal{L}_{\text{HCL}}$ in Eq. \ref{Eq: hierarchical contrastive loss} with two auxiliary losses above, \textit{i.e.,} prototype contrastive loss $\mathcal{L}_{\text{PCL}}$ in Eq. \ref{Eq: prototype contrastive loss} and self-distillation loss $\mathcal{L}_{\text{DL}}$ in Eq. \ref{Eq: distillation loss}, we can obtain the overall training loss $\mathcal{L}_{\text{all}}$, which is defined as:
\begin{equation}
	\mathcal{L}_{\text{all}}=\mathcal{L}_{\text{HCL}}+\lambda \mathcal{L}_{\text{DL}} + \mu \mathcal{L}_{\text{PCL}}, \label{Eq: total loss}
\end{equation}
where $\lambda$ and $\mu$ are the weighting parameters.

\textbf{Inference.} During the inference stage, we directly perform channel-wise normalization at text and video features to compute the similarity scores at three granularity levels, followed by weighting them for the final similarity matching:$S_{t,v}=S_{t,v}^{I} + \alpha S_{t,v}^{L} + \beta S_{t,v}^{G}$, $\alpha$ and $\beta$ correspond to the trade-off hyper-parameters in Eq. \ref{Eq: hierarchical contrastive loss}. 

\section{Experiments}
\label{Sec: Experiments}

\subsection{Experimental Settings}
\textbf{Datasets.} 
We carry out experiments on four widely recognized benchmarks for text-video retrieval, including: \textbf{(1) MSR-VTT} \cite{xu2016msr} contains 10,000 videos with length varying from 10s to 32s, each paired with 20 human-labeled sentences. In this paper, we adopt the widely-used 9K group as training set. The test set in both data splits is 1K group, which contains 1,000 videos following JSFusion \cite{yu2018joint}. Unless otherwise annotated, we adopt 9K as the default. \textbf{(2) ActivityNet} \cite{caba2015activitynet} consists of 20,000 YouTube videos with 100,000 captions. 
We follow \cite{luo2022clip4clip} to concatenate multiple descriptions of a video together for video-paragraph retrieval. We report the results on the ``val1'' split which contains 4,900 videos.
\textbf{(3) VATEX} \cite{wang2019vatex} includes 34,991 videos, each with multiple captions. Following the split protocol from HGR \cite{chen2020fine}, the training set, validation set and testing set contain 25,991, 1,500 and 1,500 videos, respectively. \textbf{(4) Charades} \cite{sigurdsson2016hollywood} contains 9,848 video clips, with each typically corresponding to a single caption. We adopt the same split protocol in \cite{lin2022eclipse} to validate our model.

\textbf{Evaluation Metrics.} 
For evaluation, we assess the performance of our method with the standard text-video retrieval metrics: Recall at Rank K (R@K, $\uparrow$ denotes the better), Median Rank (MdR, $\downarrow$ denotes the better) and Mean Rank (MnR, $\downarrow$ denotes the better). R@K records the percentage of correct items found in the top-K of retrieval results. Here, we set K=1, 5 and 10 in the experiments. MdR measures the median rank position of groundtruth in the ranking list, while MnR calculates the average rank position of groundtruth in the ranking list. In order to reflect the comprehensive retrieval performance in both text-to-video and video-to-text retrieval tasks, we also take the sum of all R@K as RSum. Note that the above two RSum are further summed to 
obtain SumR ($\uparrow$ denotes the better).

\textbf{Implementation Details.}
We choose HBI \cite{jin2023video} as the baseline model. Following \cite{luo2022clip4clip}, we initialize our text encoder and video encoder with pre-trained weights of CLIP-ViT-B/32 \cite{radford2021learning}. The layer number of two lightweight sequential transformer blocks is uniformly set as 4. To reduce the computational overhead, all videos are resized into 224 width or 224 height, and the frame rate is 3. We configure the feature dimension $D$ as 512. The word length $N_{w}$, frame length $N_{f}$ and batch size are 32, 12 and 128 for all datasets except ActivityNet (64 max words, 64 max frames and 64 batch size). The nearest neighbors number $K$, phrase length $N_{p}$ and clip length $N_{c}$ are uniformly set as 3, 6, and 6. We train our model with Adam optimizer for 5 epochs on MSR-VTT, VATEX, Charades while 20 epochs on ActivityNet, and adopt a cosine schedule strategy. We set the initial learning rate as 1e-7 and 1e-4 for CLIP encoders and other modules, respectively. The number of individual level prototypes $N_{I}^{p}$ and local level prototypes $N_{L}^{p}$ are set as 3 and 1. The hyper-parameters $\alpha$, $\beta$, $\lambda$, $\mu$, $\sigma$ are set as 0.5, 0.1, 0.2, 0.2, and 0.01. All experiments are conducted on 4 NVIDIA GeForce RTX 3090 24GB GPUs using PyTorch.

\subsection{Performance Comparison}
In this section, we compare our method with recent state-of-the-art works on four benchmark datasets namely MSR-VTT, ActivityNet, VATEX, and Charades. 

\begin{table*}
	\caption{\textbf{Cross-modal retrieval comparisons on the MSR-VTT dataset.} ``$\dag$'' denotes our reproduction of the methods with public code provided by the corresponding paper. ``-'' denotes the unavailable results. Bold denotes the best performance. }
	\label{table: MSR-VTT Results}
	\centering
	\normalsize
	\resizebox{\textwidth}{!}{
		\renewcommand{\arraystretch}{1.06}
		\begin{tabular}{l|cccccc|cccccc|c}
			\Xhline{1.25px}
			\multirow{2}*{Method} &  \multicolumn{6}{c|}{Text-to-Video Retrieval} & \multicolumn{6}{c|}{Video-to-Text Retrieval} & \multirow{2}*{SumR$\uparrow$} \\
			\cline{2-7} \cline{8-13} 
			& R@1$\uparrow$  & R@5$\uparrow$  & R@10$\uparrow$ & MdR$\downarrow$ &  MnR$\downarrow$ & RSum$\uparrow$
			& R@1$\uparrow$  & R@5$\uparrow$  & R@10$\uparrow$ & MdR$\downarrow$ &  MnR$\downarrow$ & RSum$\uparrow$ \\
			\hline
			\textit{\textbf{Non-CLIP}} & & & & & & & & & & & & \\
			$\textup{CE}$ \cite{liu2019use} \textcolor{gray}{\textit{\textbf{[BMVC'19]}}} & 20.9 & 48.8 & 62.4 & 6.0 &  28.2 &  132.1 & 20.6 & 50.3 & 64.0 & 5.3 & 25.1 & 134.9 & 267.0 \\
			$\textup{MMT}$ \cite{gabeur2020multi} \textcolor{gray}{\textit{\textbf{[ECCV'20]}}} & 26.6 & 57.1 & 69.6 &  4.0 & 24.0 & 153.3 & 27.0 & 57.5 & 69.7 & 3.7 & 21.3 & 154.2 & 307.5 \\
			$\textup{HiT}$ \cite{liu2021hit} \textcolor{gray}{\textit{\textbf{[ICCV'21]}}} & 30.7 & 60.9 & 73.2 & 2.6 & – & 164.8 & 32.1 & 62.7 & 74.1 & 3.0 & – & 168.9 & 333.7 \\
			$\textup{ClipBERT}$ \cite{lei2021less} \textcolor{gray}{\textit{\textbf{[CVPR'21]}}} & 22.0 & 46.8 & 59.9 & 6.0 & - & 128.7 & - & - & - & - & - & - & - \\
			$\textup{Frozen}$ \cite{bain2021frozen} \textcolor{gray}{\textit{\textbf{[ICCV'21]}}} & 31.0 & 59.5 & 70.5 & 3.0 & – & 161.0 & - & - & - & - & - & - & - \\
			$\textup{MCQ}$ \cite{ge2022bridging} \textcolor{gray}{\textit{\textbf{[CVPR'22]}}} & 37.6 & 64.8 & 75.1 & 3.0 & - & 177.5 & - & - & - & - & - & - & - \\
			\hline
			\hline
			\textit{\textbf{ViT-B/32}} & & & & & & & & & & & & \\
			$\textup{CLIP4Clip}$ \cite{luo2022clip4clip} \textcolor{gray}{\textit{\textbf{[Neurocomputing'22]}}} & 44.5 & 71.4 & 81.6 & \textbf{2.0} & 15.3 & 197.5 & 42.7 & 70.9 & 80.6 & \textbf{2.0} & 11.6 & 194.2 & 391.7 \\
			CenterCLIP \cite{zhao2022centerclip} \textcolor{gray}{\textit{\textbf{[SIGIR'22]}}} & 44.2 & 71.6 & 82.1 & \textbf{2.0} & 15.1 & 197.9 & 42.8 & 71.7 & 82.2 & \textbf{2.0} & 10.9 & 196.7 & 394.6  \\
			$\textup{X-Pool}$ \cite{gorti2022x} \textcolor{gray}{\textit{\textbf{[CVPR'22]}}}  & 46.9 & 72.8 & 82.2 & \textbf{2.0} & 14.3 & 201.9 & 44.4 & 73.3 & 84.0 & \textbf{2.0} & 9.0 & 201.7 & 403.6 \\
			$\textup{X-CLIP}$ \cite{ma2022x} \textcolor{gray}{\textit{\textbf{[ACM MM'22]}}} & 46.1 & 73.0 & 83.1 & \textbf{2.0} & 13.2 & 202.2 & 46.8 & 73.3 & 84.0 & \textbf{2.0} & 9.1 & 204.1 & 406.3 \\
			$\textup{EMCL-Net}$ \cite{jin2022expectation} \textcolor{gray}{\textit{\textbf{[NeurIPS'22]}}} & 46.8 & 73.1 & 83.1 & \textbf{2.0} & - & 203.0 & 46.5 & 73.5 & 83.5 & \textbf{2.0} & - & 203.5 & 406.5 \\
			$\textup{TS2-Net}$ \cite{liu2022ts2} \textcolor{gray}{\textit{\textbf{[ECCV'22]}}} & 47.0 & 74.5 & 83.8 & \textbf{2.0} & 13.0 & 205.3 & 45.3 & 74.1 & 83.7 & \textbf{2.0} & 9.2 & 203.1 & 408.4 \\
			$\textup{DRL}$ \cite{wang2022disentangled} \textcolor{gray}{\textit{\textbf{[Arxiv'22]}}} & 47.4 & 74.6 & 83.8 & \textbf{2.0} & - & 205.8 & 45.3 & 73.9 & 83.3 & \textbf{2.0} & - & 202.5 & 408.3 \\
			$\textup{DiffusionRet}$ \cite{jin2023diffusionret} \textcolor{gray}{\textit{\textbf{[ICCV'23]}}} & 49.0 & 75.2 & 82.7 & \textbf{2.0} & \textbf{12.1} & 206.9 & \textbf{47.7} & 73.8 & \textbf{84.5} & \textbf{2.0} & 8.8 & \textbf{206.0} & 412.9 \\
			$\textup{UCoFiA}$ \cite{wang2023unified} \textcolor{gray}{\textit{\textbf{[ICCV'23]}}}  & \textbf{49.4} & 72.1 & 83.5 & \textbf{2.0} & 12.9 & 205.0 & 47.1 & 74.3 & 83.0 & \textbf{2.0} & - & 204.4 & 409.4 \\
			$\textup{ProST}$ \cite{li2023progressive} \textcolor{gray}{\textit{\textbf{[ICCV'23]}}} & 48.2 & 74.6 & 83.4 & \textbf{2.0} & 12.4 & 206.2 & 46.3 & 74.2 & 83.2 & \textbf{2.0} & 8.7 & 203.7 & 409.9 \\
			EERCF \cite{tian2024towards} \textcolor{gray}{\textit{\textbf{[AAAI'24]}}} & 47.8 & 74.1 & 84.1 & - & - & 206.0 & 44.7 & 74.2 & 83.9 & - & - & 202.8 & 408.8 \\
			TeachCLIP \cite{tian2024holistic} \textcolor{gray}{\textit{\textbf{[CVPR'24]}}} & 46.8 & 74.3 & 82.6 & - & - & 203.7 & - & - & - & - & - & - & - \\
			TC-MGC \cite{jing2025tc} \textcolor{gray}{\textit{\textbf{[Information Fusion'25]}}} & 47.4 & 74.8 & \textbf{84.2} & \textbf{2.0} & 12.8 & 206.4 & 45.9 & 74.5 & 83.3 & \textbf{2.0} & 8.6 & 203.7 & 410.1 \\
			S2CA \cite{li2025s2ca} \textcolor{gray}{\textit{\textbf{[Neurocomputing'25]}}} & 47.7 & 73.4 & 83.1 & - & - & 204.2 & 45.8 & 73.4 & 83.1 & - & - & 202.3 & 406.5 \\
			$\textup{HBI}^{\dag}$ \cite{jin2023video} \textcolor{gray}{\textit{\textbf{[CVPR'23]}}} & 47.7 & 73.2 & 82.5 & \textbf{2.0} & 12.9 & 203.4 & 45.3 & 73.4 & 83.2 & \textbf{2.0} & 9.0 & 201.9 & 405.3 \\
			\rowcolor{blue!5}
			\textbf{PHA-Net (Ours)} & 48.1 & \textbf{76.4} & 84.1 & \textbf{2.0} & \textbf{12.1} & \textbf{208.6} & 46.5 & \textbf{75.1} & 83.9 & \textbf{2.0} & \textbf{8.5} & 205.5 & \textbf{414.1} \\
			\hline
			$\textup{HBI}^{\dag}$ \cite{jin2023video} + DSL \cite{cheng2021improving} & 46.9 & 73.7 & 82.4 & \textbf{2.0} & 12.7 & 203.0 & 45.7 & 73.9 & \textbf{84.1} & 2.0 & \textbf{8.8} & 203.7 & 406.7 \\
			\rowcolor{blue!5}
			\textbf{$\textup{PHA-Net}$ (Ours) + DSL \cite{cheng2021improving}} & \textbf{49.8} & \textbf{77.4} & \textbf{86.4} & \textbf{2.0} & \textbf{10.4} & \textbf{213.6} & \textbf{51.0} & \textbf{75.2} & 82.9 & \textbf{1.0} & 9.2 & \textbf{209.1}& \textbf{422.7} \\
			\hline
			\hline
			\textit{\textbf{ViT-B/16}} & & & & & & & & & & & & \\
			$\textup{CLIP4Clip}^{\dag}$ \cite{luo2022clip4clip} \textcolor{gray}{\textit{\textbf{[Neurocomputing'22]}}} & 46.4 & 72.1 & 82.0 & \textbf{2.0} & 14.7 & 200.5 & 45.4 & 73.4 & 82.4 & \textbf{2.0} & 10.7 & 201.2 & 401.7 \\
			CenterCLIP \cite{zhao2022centerclip} \textcolor{gray}{\textit{\textbf{[SIGIR'22]}}} & 48.4 & 73.8 & 82.0 & \textbf{2.0} & 13.8 & 204.2 & 47.7 & 75.0 & 83.3 & \textbf{2.0} & 10.2 & 206.0 & 410.2 \\
			$\textup{X-Pool}^{\dag}$ \cite{gorti2022x} \textcolor{gray}{\textit{\textbf{[CVPR'22]}}} & \textbf{49.7} & 74.7 & 84.2 & \textbf{2.0} & 12.3 & 208.6 & 48.1 & 76.0 & 85.5 & \textbf{2.0} & 8.1 & 209.6 & 418.2 \\
			$\textup{X-CLIP}^{\dag}$ \cite{ma2022x} \textcolor{gray}{\textit{\textbf{[ACM MM'22]}}} & 49.4 & 75.7 & 84.4 & \textbf{2.0} & 12.2 & 209.5 & 48.6 & 75.2 & 84.6 & \textbf{2.0} & 8.0 & 208.4 & 417.9 \\
			$\textup{DRL}^{\dag}$ \cite{wang2022disentangled} \textcolor{gray}{\textit{\textbf{[Arxiv'22]}}} & 49.4 & 76.4 & 84.2 & \textbf{2.0} & 13.2 & 210.0 & 47.0 & 77.1 & 84.4 & \textbf{2.0} & 9.2 & 208.5 & 418.5 \\
			$\textup{UCoFiA}^{\dag}$ \cite{wang2023unified} \textcolor{gray}{\textit{\textbf{[ICCV'23]}}} & \textbf{49.7} & 75.7 & 84.2 & \textbf{2.0} & 12.6 & 209.6 & 48.1 & 76.3 & 84.4 & \textbf{2.0} & 8.8 & 208.8 & 418.4 \\
			$\textup{ProST}^{\dag}$ \cite{li2023progressive} \textcolor{gray}{\textit{\textbf{[ICCV'23]}}} & 46.4 & 74.1 & 84.5 & \textbf{2.0} & \textbf{12.0} & 205.0 & 47.8 & 75.0 & 84.6 & \textbf{2.0} & 8.7 & 207.4 & 412.4 \\
			TC-MGC \cite{jing2025tc} \textcolor{gray}{\textit{\textbf{[Information Fusion'25]}}} & 49.0 & 75.7 & 85.4 & \textbf{2.0} & 13.2 & 210.1 & 46.4 & 77.1 & 85.3 & \textbf{2.0} & 8.8 & 208.8 & 418.9 \\
			$\textup{HBI}^{\dag}$  \cite{jin2023video} \textcolor{gray}{\textit{\textbf{[CVPR'23]}}} & 49.1 & 76.3 & \textbf{85.9} & \textbf{2.0} & 12.4 & 211.3 & 48.3 & 76.4 & \textbf{86.6} & \textbf{2.0} & 7.8 & 211.3 & 422.6 \\
			\rowcolor{blue!5}
			\textbf{PHA-Net (Ours)} & 49.6 & \textbf{76.7} & 85.2 & \textbf{2.0} & 12.6 & \textbf{211.5} & \textbf{49.4} & \textbf{77.4} & 85.1 & \textbf{2.0} & \textbf{7.7} & \textbf{211.9} & \textbf{423.4} \\
			\hline
			$\textup{HBI}^{\dag}$ \cite{jin2023video} + DSL \cite{cheng2021improving} & 49.8 & 76.7 & 86.0 & 2.0 & 11.8 & 212.5 & 49.8 & \textbf{78.0} & \textbf{86.7} & 2.0 & 7.7 & 214.5 & 427.0 \\
			\rowcolor{blue!5}
			\textbf{$\textup{PHA-Net}$ (Ours) + DSL \cite{cheng2021improving}} & \textbf{52.4} & \textbf{77.6} & \textbf{87.1} & \textbf{1.0} & \textbf{10.5} & \textbf{217.1} & \textbf{54.6} & 77.6 & 85.6 & \textbf{1.0} & \textbf{7.5} & \textbf{217.8} & \textbf{434.9} \\
			
			\Xhline{1.25px}
		\end{tabular}
	}
\end{table*}

\begin{table*}
	\caption{\textbf{Cross-modal retrieval comparisons on other datasets.} The left section details the results from the corresponding paper, while the middle and right sections delineate the reproduced results. ``-'' denotes the unavailable results. Bold denotes the best performance.}
	\label{table: Other Results}
	\centering
	\huge
	\resizebox{\textwidth}{!}{
		\renewcommand{\arraystretch}{1.06}
		\begin{tabular}{l|ccccc|ccccc|ccccc}
			\Xhline{2.5px}
			\multirow{2}*{Method} & \multicolumn{5}{c|}{ActivityNet} & \multicolumn{5}{c|}{VATEX} & \multicolumn{5}{c}{Charades}\\
			\cline{2-6} \cline{7-11} \cline{12-16}
			& R@1$\uparrow$  & R@5$\uparrow$  & R@10$\uparrow$ & MnR$\downarrow$ &  RSum$\uparrow$
			& R@1$\uparrow$  & R@5$\uparrow$  & R@10$\uparrow$ & MnR$\downarrow$ &  RSum$\uparrow$
			& R@1$\uparrow$  & R@5$\uparrow$  & R@10$\uparrow$ & MnR$\downarrow$ &  RSum$\uparrow$\\
			\hline
			\textit{\textbf{Text-to-Video Retrieval}} & & & & & & & & & & & & \\
			$\textup{HGR}$ \cite{chen2020fine} \textcolor{gray}{\textit{\textbf{[CVPR'20]}}} & - & - & - & - & - & 35.1 & 73.5 & 83.5 & - & 192.1 & - & - & - & - & - \\
			$\textup{SupportSet}$ \cite{patrick2020support} \textcolor{gray}{\textit{\textbf{[ICLR'21]}}} & 29.2 & 61.6 & 94.7 & - & 185.5 & 45.9 & 82.4 & 90.4 & - & 218.7 & - & - & - & - & - \\
			$\textup{CLIP4Clip}$ \cite{luo2022clip4clip} \textcolor{gray}{\textit{\textbf{[Neurocomputing'22]}}} & 40.5 & 72.4 & 83.6 & 7.5 & 196.5 & 58.7 & 89.3 & 94.7 & 3.7 & 242.7 & 11.8 & 28.1 & 37.4 & 94.7 & 77.3 \\
			$\textup{TS2-Net}$ \cite{liu2022ts2} \textcolor{gray}{\textit{\textbf{[ECCV'22]}}} & 41.0 & 73.6 & 84.5 & 8.4 & 199.1 & 59.5 & 89.8 & 95.0 & 3.6 & 244.3 & 10.4 & 26.5 & 36.6 & 89.8 & 73.5\\
			$\textup{X-CLIP}$ \cite{ma2022x} \textcolor{gray}{\textit{\textbf{[ACM MM'22]}}} & 44.3 & 74.1 & - & 7.9 & - & 59.1 & 88.9 & 94.2 & 3.9 & 242.2 & 11.4 & 27.9 & 37.4 & 93.2 & 76.7\\
			$\textup{EMCL-Net}$ \cite{jin2022expectation} \textcolor{gray}{\textit{\textbf{[NeurIPS'22]}}} & 41.2 & 72.7 & - & - & - & 59.1 & 88.8 & 94.3 & 3.9 & 242.2 & 10.6 & 24.7 & 33.5 & 97.5 & 68.8 \\
			$\textup{UATVR}$ \cite{fang2023uatvr} \textcolor{gray}{\textit{\textbf{[ICCV'23]}}} & - & - & - & - & - & 58.8 & 89.6 & 94.7 & 3.7 & 243.1 & 10.5 & 26.6 & 36.7 & 95.5 & 73.8\\
			$\textup{ProST}$ \cite{li2023progressive} \textcolor{gray}{\textit{\textbf{[ICCV'23]}}} & - & - & - & - & - & 57.5 & 89.3 & 94.7 & 3.9 & 241.5 & 9.1 & 24.0 & 32.8 & 98.0 & 65.9 \\
			$\textup{DiffusionRet}$ \cite{jin2023diffusionret} \textcolor{gray}{\textit{\textbf{[ICCV'23]}}} & 45.8 & 75.6 & 86.3 & 6.5 & 207.7 & 60.6 & 90.3 & 95.1 & \textbf{3.5} & 246.0 & 12.3 & 30.4 & 40.3 & 90.3 & 83.0 \\
			$\textup{HBI}$  \cite{jin2023video} \textcolor{gray}{\textit{\textbf{[CVPR'23]}}} & 42.2 & 73.0 & 84.6 & 6.6 & 199.8 & 60.3 & \textbf{90.4} & 95.2 & 3.6 & 245.9 & 13.0 & 31.8 & 41.2 & 86.1 & 86.0\\
			\rowcolor{blue!5}
			\textbf{PHA-Net (Ours)} & \textbf{46.2} & \textbf{76.9} & \textbf{87.3} & \textbf{6.3} & \textbf{210.4} & \textbf{60.7} & 90.2 & \textbf{95.3} & \textbf{3.5} & \textbf{246.2} & \textbf{13.6} & \textbf{32.5} & \textbf{41.8} & \textbf{81.7} & \textbf{87.9} \\
			\hline
			\hline
			\textit{\textbf{Video-to-Text Retrieval}} & & & & & & & & & & & & \\
			$\textup{SupportSet}$ \cite{patrick2020support} \textcolor{gray}{\textit{\textbf{[ICLR'21]}}} & 28.7 & 60.8 & 94.8 & - & 184.3 & 61.2 & 85.2 & 91.8 & - & 238.2 & - & - & - & - & - \\
			$\textup{CLIP4Clip}$ \cite{luo2022clip4clip} \textcolor{gray}{\textit{\textbf{[Neurocomputing'22]}}} & 41.4 & 73.7 & 85.3 & 6.7 & 200.4 & 74.8 & 96.7 & 98.6 & 1.8 & 270.1 & 10.8 & 29.7 & 38.2 & 94.8 & 78.7 \\
			$\textup{TS2-Net}$ \cite{liu2022ts2} \textcolor{gray}{\textit{\textbf{[ECCV'22]}}} & - & - & - & - & - & 74.6 & 96.3 & 98.9 & 1.8 & 269.8 & 10.5 & 27.5 & 37.6 & 94.5 & 75.6\\
			$\textup{X-CLIP}$ \cite{ma2022x} \textcolor{gray}{\textit{\textbf{[ACM MM'22]}}} & 43.9 & 73.9 & - & 7.6 & - & 74.8 & \textbf{97.3} & 99.0 & 1.8 & 271.1 & 11.3 & 28.3 & 38.1 & 92.4 & 77.7\\
			$\textup{EMCL-Net}$ \cite{jin2022expectation} \textcolor{gray}{\textit{\textbf{[NeurIPS'22]}}} & 42.7 & 74.0 & - & - & - & 74.4 & 96.8 & 98.5 & 1.9 & 269.7 & 9.7 & 24.2 & 33.2 & 102.8 & 67.1 \\
			$\textup{UATVR}$ \cite{fang2023uatvr} \textcolor{gray}{\textit{\textbf{[ICCV'23]}}} & - &  - & - & - & - & 75.7 & 96.7 & 98.9 & 1.9 & 271.3 & 10.7 & 27.6 & 37.5 & 93.8 & 75.8 \\
			$\textup{ProST}$ \cite{li2023progressive} \textcolor{gray}{\textit{\textbf{[ICCV'23]}}} & - & - & - & - & - & 74.3 & 96.6 & 98.7 & 1.8 & 269.6 & 9.5 & 24.9 & 35.1 & 98.9 & 69.5 \\
			$\textup{DiffusionRet}$ \cite{jin2023diffusionret} \textcolor{gray}{\textit{\textbf{[ICCV'23]}}} & - & - & - & - & - & 75.5 & 97.2 & 98.9 & \textbf{1.7} & \textbf{271.6} & 13.1 & 31.2 & 40.6 & 92.9 & 84.9 \\
			$\textup{HBI}$ \cite{jin2023video} \textcolor{gray}{\textit{\textbf{[CVPR'23]}}} & 42.4 & 73.0 & 86.0 & 6.5 & 201.4 & 75.0 & 97.1 & \textbf{99.1} & \textbf{1.7} & 271.2 & 13.3 & 31.6 & 41.9 & 88.4 & 86.8 \\
			\rowcolor{blue!5}
			\textbf{PHA-Net (Ours)} & \textbf{45.1} & \textbf{77.1} & \textbf{87.8} & \textbf{6.2} & \textbf{210.0} & \textbf{75.9} & 96.8 & 98.9 & \textbf{1.7} & \textbf{271.6} & \textbf{13.6} & \textbf{32.5} & \textbf{43.7} & \textbf{81.9} & \textbf{89.8} \\
			\Xhline{2.5px}
		\end{tabular}
	}
\end{table*}

\textbf{Compared methods.} The details of compared methods are listed as follows:\par
$\bullet$ CE \cite{liu2019use} uses a collaborative gating mechanism to fuse multiple modal features into a compact video representation. 

$\bullet$ MMT \cite{gabeur2020multi} aggregates feature extracted by different experts via a cross-modal encoder.

$\bullet$ HiT \cite{liu2021hit} performs hierarchical cross-modal contrastive matching with momentum contrast for text-video retrieval.

$\bullet$ ClipBERT \cite{lei2021less} employs sparse sampling mechanism to enable affordable end-to-end video-and-language learning.

$\bullet$ Frozen \cite{bain2021frozen} introduces a dual encoder model for efficient text-video retrieval.

$\bullet$ MCQ \cite{ge2022bridging} uses a pretext task to enable fine-grained interactions and maintain high efficiency for video retrieval.

$\bullet$ HGR \cite{chen2020fine} decomposes video-text matching into global-to-local levels for hierarchical graph reasoning.

$\bullet$ SupportSet \cite{patrick2020support} leverages a generative model to naturally push related samples together.

$\bullet$ CLIP4Clip \cite{luo2022clip4clip} transfers CLIP knowledge to text-video retrieval and investigates three similarity calculators for global semantic alignment.

$\bullet$ X-Pool \cite{gorti2022x} attempts to generate an aggregated video representation conditioned on text's attention weights over the frames.

$\bullet$ CenterCLIP \cite{zhao2022centerclip} devises a multi-segment token cluster module to reduce the number of redundant video tokens.

$\bullet$ TS2-Net \cite{liu2022ts2} presents a novel token shift and selection transformer architecture to capture subtle movements and enhance salient object modeling ability.

$\bullet$ X-CLIP \cite{ma2022x} leverages cross-grained contrasts to mitigate the negative effects of unnecessary words and frames.

$\bullet$ EMCL-Net \cite{jin2022expectation} innovatively learns compact video-and-language representation in an expectation-maximization contrastive learning manner.

$\bullet$ DRL \cite{wang2022disentangled} proposes a weighted token-wise interaction and channel decorrelation regularization for pair-wise correlations exploration and feature redundancy decrease.

$\bullet$ UATVR \cite{fang2023uatvr} introduces semantic aggregation learnable tokens and probabilistic embeddings to regard each text-video lookup as a distribution matching process.

$\bullet$ DiffusionRet \cite{jin2023diffusionret} creatively uses the diffusion model to model the the joint probability distribution of text and video for cross-modal retrieval task.

$\bullet$ UCoFiA \cite{wang2023unified} simultaneously considers cross-modal correspondence from different granularity and multi-grained alignments in a unified retrieval model.

$\bullet$ ProST \cite{li2023progressive} decomposes the matching process into complementary object-phrase and event-sentence prototype alignments to achieve progressive spatio-temporal prototype matching.     

$\bullet$ EERCF \cite{tian2024towards} devises a text-gated interaction block and a combination of inter-feature and intra-feature loss function for efficient and effective text-video retrieval.

$\bullet$ TeachCLIP \cite{tian2024holistic} enables a CLIP4Clip-based network to learn from more advanced yet computationally heavy TVR models.

$\bullet$ TC-MGC \cite{jing2025tc} aggregates frame features into semantic-relevant video and frame representations in sentence-guided and word-guided manners.

$\bullet$ S2CA \cite{li2025s2ca} decouples texts and videos into several conceptual aggregated prototypes to achieve concept-level alignment.

$\bullet$ HBI \cite{jin2023video} models video-language learning as a multivariate cooperative game and uses a hierarchical banzhaf interaction to value relevance between frames and words.

\textbf{Results on MSR-VTT.} The comparison results on the MSR-VTT dataset are shown in Table \ref{table: MSR-VTT Results}. Overall, our model significantly outperforms existing methods on most of the evaluation metrics, with the best result being SumR=414.1 for the overall performance. Compared to the baseline HBI, PHA-Net improves the R@1 metric from 47.7 to 48.1 (\textit{t2v}) and from 45.3 to 46.5 (\textit{v2t}). Since HBI also conducts hierarchical alignments, we owe the performance gain to three core designs, including the modality-shared prototypes, prototype-supported token merge module, and prototype contrastive loss. Even compared with the state-of-the-art methods, \textit{i.e.,} EERCF, TeachCLIP, and TC-MGC,  PHA-Net still surpasses them by 2.6\%, 4.9\%, and 2.2\% in \textit{t2v} RSum. These results highlight the benefits of shared prototypes in facilitating cross-modal alignment. We also compare PHA-Net to the recent prototype-related model S2CA and find that PHA-Net shows clear performance improvements over S2CA across all recall metrics. Although S2CA uses modality-shared prototypes to model fine-grained concept association, this model only performs a single concept-level alignment without considering hierarchical correspondences between texts and videos, where our PHA-Net fills this blank and achieves better performance. When DSL \cite{cheng2021improving} strategy is used in the inference stage, our model still outperforms the baseline by 16.0\% on SumR. Besides, after being equipped with a stronger ViT-B/16 backbone, the SumR of our model can be further improved to 423.4, which shows 0.8\% performance gain over the baseline. Similarly, the utilization of DSL \cite{cheng2021improving} strategy also improves the SumR of our model by 7.9\% from HBI. The above results clearly confirm the effectiveness of our method.

\textbf{Results on Other Datasets.} To validate the robustness of our method, we provide quantitative results on the ActivityNet, VATEX, and Charades datasets in Table \ref{table: Other Results}. All models employ only CLIP-ViT-B/32 without any post-processing operations. By observation, our PHA-Net shows consistent performance improvements across versatile datasets. For example, PHA-Net surpasses other compared methods on all metrics by a large margin on the ActivityNet dataset, with the best results being 46.2 R@1 in \textit{t2v} and 45.1 R@1 in \textit{v2t}. We think the main reason is that ActivityNet contains more complex and long-term movie videos, and the introduced modality-shared prototypes enable the model to handle intricate textual-visual relationships and achieve accurate matching with these videos. For the VATEX dataset, our PHA-Net brings 0.4\% and 0.9\% performance gains on the R@1 metric in comparison with the baseline method, demonstrating the importance of prototype-guided alignment for hierarchical text-video retrieval. We also find that PHA-Net also improves the recent competitors, such as UATVR and ProST, by 3.1\% and 4.7\% at \textit{t2v} RSum. The results verify the above thought. Meanwhile, compared to the RSum metrics of the baseline method on the Charades dataset, our PHA-Net achieves comparable results with 1.9\% and 3.0\% absolute performance improvements on \textit{t2v} and \textit{v2t} splits, exhibiting the generalization capabilities of our model.

\subsection{Ablation Study}

\textbf{Hierarchical Alignment.} As mentioned above, cross-modal alignments at individual, local, and global level are complementary to each other for text-video retrieval. As presented in Table \ref{table: hierarchical structure ablation}, individual+local alignments and individual+global alignments surpass individual-only alignment in R@5 (\textit{e.g.,} 75.8 \textit{v.s.} 75.4 and 74.9 \textit{v.s.} 74.5 in \textit{t2v}, 75.7 \textit{v.s.} 75.4 and 74.8 \textit{v.s.} 74.5 in \textit{v2t}), which shows the effectiveness of local-level alignment and global-level alignment. Meanwhile, we notice that employing hierarchical alignment at all three levels can further improve the performance, with the best results being R@5=76.4 for text-to-video retrieval and R@5=75.1 for video-to-text retrieval. These results justify that
the three levels are complementary and hierarchical alignment at them is critical to capture rich semantics in video and corresponding text.
 
\begin{table}
	\caption{\textbf{Ablation study of hierarchical alignment at different levels. }}
	\label{table: hierarchical structure ablation}
	\centering
	\normalsize
	\resizebox{0.5\textwidth}{!}{
		\renewcommand{\arraystretch}{1.05} 
		\begin{tabular}{l|ccc|ccc}
			\Xhline{1.25px}
			\multirow{2}*{Hierarchical} & \multicolumn{3}{c|}{Tex-to-Video} & \multicolumn{3}{c}{Video-to-Text} \\
			\cline{2-4} \cline{5-7} 
			& R@1$\uparrow$  & R@5$\uparrow$  & R@10$\uparrow$ 
			& R@1$\uparrow$  & R@5$\uparrow$  & R@10$\uparrow$ \\
			\hline
			Individual-only & 47.8 & 75.4 & 83.7 & \textbf{47.1} & 74.5 & 82.9 \\
			Individual + Local & 47.8 & 75.8 & 83.9 & 46.6 & 74.9 & 83.7 \\
			Individual + Global & 47.6 & 75.7 & 83.9 & \textbf{47.1} & 74.8 & 83.1 \\
			\rowcolor{blue!5}
			All & \textbf{48.1} & \textbf{76.4} & \textbf{84.1} & 46.5 & \textbf{75.1} & \textbf{83.9} \\
			\Xhline{1.25px}
		\end{tabular}
	}
\end{table}

\textbf{Shared Strategy.} In Table \ref{table: shared strategy ablation}, we evaluate the influence of the shared strategy for learnable prototypes (LP) and transformer blocks (TB) at individual and local levels. Overall, our method with the shared strategy for LP or TB consistently achieves better \textit{t2v} R@1 results than the unshared strategy. We explain it is because the shared strategy can implicitly facilitate the interaction between hierarchical textual and visual feature, so it is more suitable to bridge the modality gap between them. It is worth noting that compared to shared LP or TB at the local level, applying the shared strategy at the individual level gets a further performance boost at \textit{v2t} R@1 (\textit{i.e.,} 46.4 \textit{v.s.} 46.1 for LP, 46.5 \textit{v.s.} 45.6 for TB). The possible reason is that individual-level alignment is crucial for bridging the semantic gap between text and video data. Additionally, we find that using a shared strategy makes the model more efficient with lower inference memory usage and smaller trainable parameters, which simultaneously ensure the effectiveness and efficiency of our approach. Therefore, we choose this strategy by default for our experiments.

\begin{table}
	\caption{\textbf{Ablation study of shared strategy for learnable prototypes (LP) and transformer blocks (TB) at individual (ILP/ITB) and local (LLP/LTB) levels.} \textcolor{gray}{\(\clubsuit\)} and \textcolor{blue}{\(\clubsuit\)} denote unshared and shared learning strategies, respectively. ``Inference Memory'' denotes the memory usage in the inference stage.  ``\#Param.'' denotes the number of trainable parameters.}
	\label{table: shared strategy ablation}
	\centering
	\huge
	\resizebox{0.5\textwidth}{!}{
		\renewcommand{\arraystretch}{1.05} 
		\begin{tabular}{cc|ccc|ccc|c}
			\Xhline{2px}
			\multirow{2}*{ILP} & \multirow{2}*{LLP} & \multicolumn{3}{c|}{Text-to-Video} & \multicolumn{3}{c|}{Video-to-Text} & \multirow{2}*{\makecell[c]{Inference\\Memory$\downarrow$}} \\
			\cline{3-5} \cline{6-8} 
			&& R@1$\uparrow$  & R@5$\uparrow$  & R@10$\uparrow$
			& R@1$\uparrow$  & R@5$\uparrow$  & R@10$\uparrow$ \\
			\hline
			\textcolor{gray}{\(\clubsuit\)} & \textcolor{gray}{\(\clubsuit\)} & 46.2 & 73.0 & 83.4 & 46.0 & 72.6 & 83.3 & 3044.356MB\\
			\textcolor{blue}{\(\clubsuit\)} & \textcolor{gray}{\(\clubsuit\)} & 46.5 & 73.0 & 83.7 & 46.4 & 73.9 & \textbf{83.9} & 3044.350MB\\
			\textcolor{gray}{\(\clubsuit\)} & \textcolor{blue}{\(\clubsuit\)} & 47.8 & 74.8 & 83.5 & 46.1 & 73.7 & 83.4 & 3044.354MB \\
			\rowcolor{blue!5}
			\textcolor{blue}{\(\clubsuit\)} & \textcolor{blue}{\(\clubsuit\)} & \textbf{48.1} & \textbf{76.4} & 84.1 & \textbf{46.5} & \textbf{75.1} & \textbf{83.9} & \textbf{3044.348MB} \\
			\hline
			\hline
			
			\multirow{2}*{ITB} & \multirow{2}*{LTB} & \multicolumn{3}{c|}{Text-to-Video} & \multicolumn{3}{c|}{Video-to-Text} & \multirow{2}*{\#Param.$\downarrow$} \\
			\cline{3-5} \cline{6-8} 
			&& R@1$\uparrow$  & R@5$\uparrow$  & R@10$\uparrow$
			& R@1$\uparrow$  & R@5$\uparrow$  & R@10$\uparrow$ \\
			\hline
			\textcolor{gray}{\(\clubsuit\)} & \textcolor{gray}{\(\clubsuit\)} & 45.7 & 74.2 & 83.3 & \textbf{46.5} & 72.8 & 82.2 & 129.39M \\
			\textcolor{blue}{\(\clubsuit\)} & \textcolor{gray}{\(\clubsuit\)} & 46.5 & 72.9 & 83.6 & \textbf{46.5} & 73.9 & 83.8 & 120.99M \\
			\textcolor{gray}{\(\clubsuit\)} & \textcolor{blue}{\(\clubsuit\)} & 46.3 & 74.0 & \textbf{84.5} & 45.6 & 73.0 & 81.9 & 120.99M \\
			\rowcolor{blue!5}
			\textcolor{blue}{\(\clubsuit\)} & \textcolor{blue}{\(\clubsuit\)} & \textbf{48.1} & \textbf{76.4} & 84.1 & \textbf{46.5} & \textbf{75.1} & \textbf{83.9} & \textbf{112.59M} \\
			\Xhline{2px}
		\end{tabular}
	}
\end{table}

\textbf{Learning Objective.} In Table \ref{table: learning objectives ablation}, we study the impact of different learning objectives for PHA-Net with four control groups: (i) baseline only with $\mathcal{L}_{\text{HCL}}$; (ii) adding $\mathcal{L}_{\text{PCL}}$ into (i); (iii) adding $\mathcal{L}_{\text{DL}}$ into (i); and (iv) full model that integrates all of the aforementioned learning objectives. Compared with baseline (i), incorporating $\mathcal{L}_{\text{PCL}}$ improves the R@1 and R@10 metrics from 46.9 to 47.5 (\textit{t2v}) and from 84.0 to 84.7 (\textit{t2v}), verifying the prototype diversity optimization is crucial to text-video matching. Next, incorporating $\mathcal{L}_{\text{DL}}$ into the baseline (i) brings a further performance enhancement, with 2.1\% and 0.3\% absolute performance gains on the R@5 metric in text-to-video retrieval and video-to-text retrieval. This indicates the advantage of self-distillation loss in improving the generalization ability. Finally, combining all the above learning objectives (iv) yields the best performance. 

\begin{table}
    \caption{\textbf{Ablation study of learning objective in PHA-Net.} $\mathcal{L}_{\text{HCL}}$, $\mathcal{L}_{\text{PCL}}$, and $\mathcal{L}_{\text{DL}}$ represent hierarchical contrastive loss, prototype contrastive loss and self-distillation loss, respectively. }
    \label{table: learning objectives ablation}
    \centering
    \huge
    \resizebox{0.5\textwidth}{!}{
        \renewcommand{\arraystretch}{1} 
        \begin{tabular}{ccc|ccc|ccc}
            \Xhline{2.5px}
            \multirow{2}*{$\mathcal{L}_{\text{HCL}}$} &  \multirow{2}*{$\mathcal{L}_{\text{PCL}}$} &  \multirow{2}*{$\mathcal{L}_{\text{DL}}$} & \multicolumn{3}{c|}{Text-to-Video} & \multicolumn{3}{c}{Video-to-Text} \\
            \cline{4-6} \cline{7-9} 
            &&&  R@1$\uparrow$  & R@5$\uparrow$ & R@10$\uparrow$
            & R@1$\uparrow$  & R@5$\uparrow$ & R@10$\uparrow$ \\
            \hline
            \Checkmark & \textcolor{gray}{\XSolidBrush} & \textcolor{gray}{\XSolidBrush} & 46.9 & 74.0 & 84.0 & \textbf{46.8} & 74.7 & 84.6 \\
            \Checkmark & \Checkmark & \textcolor{gray}{\XSolidBrush} & 47.5 & 74.0 & \textbf{84.7} & 46.4 & 74.3 & \textbf{84.7} \\
            \Checkmark & \textcolor{gray}{\XSolidBrush} & \Checkmark & 46.9 & 76.1 & 83.8 & 45.5 & 75.0 & 83.8 \\
            \rowcolor{blue!5}
            \Checkmark & \Checkmark & \Checkmark & \textbf{48.1} & \textbf{76.4} & 84.1 & 46.5 & \textbf{75.1} & 83.9  \\
            \Xhline{2.5px}
        \end{tabular}
    }
\end{table}

\textbf{Transformer Block.} To examine the effect of transformer block in PHA-Net, we compare the PHA-Net with and without the individual-level transformer block (ITB) as well as local-level transformer block (LTB). The experimental results are illustrated in Table \ref{table: transformer_block_ablation}. Specifically, PHA-Net without TB only achieves 45.3 \textit{t2v} R@1. However, when PHA-Net is additionally equipped with ITB or LTB, the performance achieves further absolute boost of 0.7\% and 0.5\%, which prove the advantage of TB in modeling relations between feature tokens and learnable prototypes. Particularly, it is worth noting that the usage of both ITB and LTB gets the best R@1 metric in \textit{t2v} (48.1) and \textit{v2t} (46.5). Therefore, we conclude that transformer block is a key to improving the retrieval performance.

\begin{table}
    \caption{\textbf{Ablation study of transformer block (TB) at individual level (ITB) and local level (LTB).} }
    \label{table: transformer_block_ablation}
    \centering
    \normalsize
    \resizebox{0.5\textwidth}{!}{
        \renewcommand{\arraystretch}{1.05} 
        \begin{tabular}{l|ccc|ccc}
            \Xhline{1.25px}
            \multirow{2}*{Method} & \multicolumn{3}{c|}{Text-to-Video} & \multicolumn{3}{c}{Video-to-Text} \\
            \cline{2-4} \cline{5-7} 
            & R@1$\uparrow$  & R@5$\uparrow$  & R@10$\uparrow$
            & R@1$\uparrow$  & R@5$\uparrow$  & R@10$\uparrow$ \\
            \hline
            w/o TB & 45.3 & 73.9 & 83.2 & 46.4 & 72.4 & 82.8 \\
            w/ ITB & 46.0 & 75.6 & \textbf{84.3} & 45.8 & 74.7 & 83.6 \\
            w/ LTB & 45.8 & 73.4 & 83.1 & 45.6 & 72.3 & 82.9 \\
            \rowcolor{blue!5}
            w/ ITB \& LTB & \textbf{48.1} & \textbf{76.4} & 84.1 & \textbf{46.5} & \textbf{75.1} & \textbf{83.9} \\
            \Xhline{1.25px}
        \end{tabular}
    }
\end{table}

\begin{table}
	\caption{\textbf{Ablation study of token merge module.} ``CA'', ``CCA'', and ``PCA'' denote cross-attention, center-supported cross-attention, and prototype-supported cross-attention, respectively.}
	\label{table: token merge module ablation}
	\centering
	\huge
	\resizebox{0.5\textwidth}{!}{
		\renewcommand{\arraystretch}{1.05} 
		\begin{tabular}{l|ccc|ccc|c}
			\Xhline{2.5px}
			\multirow{2}*{Method} & \multicolumn{3}{c|}{Text-to-Video} & \multicolumn{3}{c|}{Video-to-Text} & \multirow{2}*{FLOPs $\downarrow$} \\
			\cline{2-4} \cline{5-7} 
			& R@1$\uparrow$  & R@5$\uparrow$  & R@10$\uparrow$
			& R@1$\uparrow$  & R@5$\uparrow$  & R@10$\uparrow$ \\
			\hline
			w/o 1D-Conv & 47.4 & 75.0 & 83.9 & 46.1 & 73.7 & 83.5 & \textbf{36.785G} \\
			w/ 1D-Conv & \textbf{48.1} & \textbf{76.4} & \textbf{84.1} & \textbf{46.5} & \textbf{75.1} & \textbf{83.9} & 36.836G \\
			\hline
			w/o CA & 46.3 & 74.4 & 83.8 & 45.6 & 72.9 & 82.1 & \textbf{36.821G} \\
			w/ CCA \& CCA & 46.7 & 75.7 & 83.9 & 46.2 & 74.2 & 83.3 & 36.862G \\
			w/ CCA \& PCA & 47.2 & 74.0 & 84.0 & \textbf{46.6} & 73.9 & \textbf{84.5} & 36.856G \\
			w/ PCA \& CCA & 47.6 & 74.5 & 83.6 & 45.6 & 74.2 & 83.6 & 36.842G \\
			\rowcolor{blue!5}
			w/ PCA \& PCA & \textbf{48.1} & \textbf{76.4} & \textbf{84.1} & 46.5 & \textbf{75.1} & 83.9 & 36.836G \\
			\Xhline{2.5px}
		\end{tabular}
	}
\end{table}

\textbf{Token Merge Module.} As shown in the top section of the Table \ref{table: token merge module ablation}, the performance drops significantly without the one-dimensional convolutional layer. We deem the reason is that this layer can effectively enhance temporal relations between input tokens, thus obtaining significant performance gain. The results validate the above thought. Note that the increased FLOPs are acceptable compared to the clear performance enhancement. As shown in the bottom section of the Table \ref{table: token merge module ablation}, ``w/o CA'' means the clustered tokens are directly output without the cross-attention block processing. ``CCA'' means that the center-supported cross-attention, where the clustered tokens are regarded as queries $Q$, and the tokens output from the convolutional layer are regarded as keys $K$ and values $V$. ``PCA'' means our proposed prototype-supported cross-attention. The remaining four methods represent the utilization of CCA or PCA in individual-to-local and local-to-global token merging process. 

It can be found that PHA-Net with CCA or PCA performs better than that without CA, which can be ascribed to the prototype-guided or center-guided semantic-aware fusion mechanism. Compared to the combination of CCA and CCA, using PCA in the individual-to-local or local-to-global merging process achieves better \textit{t2v} R@1 results (\textit{i.e.,} 47.6 \textit{v.s.} 46.7 and 47.2 \textit{v.s.} 46.7). The main reason may be that the prototype interference will explicitly suppress the cluster noise and maintain cross-modal semantic consistency. Noting that the performance can be further improved with PCA application in all two merging process, with the best results being R@1=48.1 and R@5=76.4 in \textit{t2v}. Moreover, due to smaller length of clustered tokens, PCA has an advantage to slightly reduce computational complexity, which ensures the efficiency of our approach in practical retrieval process. Thus, we apply PCA in the token merge modules for the remaining experiments.

\textbf{Prototype Number $N_{I}^{p}$ and $N_{L}^{p}$.} To assess how many individual-level and local-level learnable prototypes the model required to learn, we evaluate the prototype number range setting $N_{I}^{p} \in \{0,1,3,5,7,9\}$ and $N_{L}^{p} \in \{0,1,3,5,7,9\}$. The Table \ref{table: prototype numbers ablation}. reveals the following key observations. First, with individual-only (\textit{i.e,} $N_{I}^{p}>0$) or local-only (\textit{i.e,} $N_{L}^{p}>0$) learnable prototypes appended, the former one achieves 0.8\% \textit{t2v} R@1 improvement (46.5 \textit{v.s.} 47.3) while the latter one suffers 1.0\% \textit{t2v} R@1 decrease (46.5 \textit{v.s.} 45.5). This is because the former explicitly facilitates local-level and global-level alignment through stacked merging modules, and the latter is insufficient to align local-level semantics without individual-level prototypes. 

Second, compared with individual-only alignment, adding one prototype at the local level brings clear performance gains in R@1 (\textit{e.g.,} 48.1 \textit{v.s.} 47.3 in \textit{t2v}, 46.5 \textit{v.s.} 45.2 in \textit{v2t}). The possible reason is that this prototype enables the model to better align local-level semantics and boost global-level alignment through the token merging process. With the increasing number of individual-level prototypes, the performance continues to grow until $N_{I}^{p}=3$, and then goes to saturation. This can be attributed to the assistance of individual-level alignment to local-level alignment. 

Third, improper configuration of individual-level and local-level prototypes may degrade performance, possibly because too many prototypes introduce noisy distractions, while too few prototypes are not enough to express semantics. Empirically, due to a significant amount of fine-grained attributes hidden in words and frames, more prototypes are required to accommodate abundant semantic information. Conversely, the merged phrases and clips are highly unified entities with homogeneous semantics. Under this circumstance, multiple prototypes may bring semantic redundancy and increase the risk of over-fitting. Therefore, we develop a pyramid-based mechanism for hierarchical prototype selection, where more prototypes are allocated to lower levels and fewer to higher ones. Based on the sensitivity analysis, the 3:1 cross-level ratio is recommended as the practical deployment guideline to achieve the optimal accuracy-efficiency trade-off. In our experiments, we set final $N_{I}^{p}=3$ and $N_{L}^{p}=1$.

\begin{table}
	\caption{\textbf{Ablation study of individual-level prototype number $N_{I}^{p}$ and local-level prototype number $N_{L}^{p}$.}}
	\label{table: prototype numbers ablation}
	\centering
	\normalsize
	\resizebox{0.5\textwidth}{!}{
		\renewcommand{\arraystretch}{1.05} 
		\begin{tabular}{c|ccc|ccc}
			\Xhline{1.2px}
			\multirow{2}*{\makecell[c]{\{$N_{I}^{p}, N_{L}^{p}$\}}} & \multicolumn{3}{c|}{Text-to-Video} & \multicolumn{3}{c}{Video-to-Text} \\
			\cline{2-4} \cline{5-7} 
			& R@1$\uparrow$  & R@5$\uparrow$  & R@10$\uparrow$
			& R@1$\uparrow$  & R@5$\uparrow$  & R@10$\uparrow$ \\
			\hline
			\{0, 0\} & 46.5 & 73.1 & 83.0 & 44.6 & 72.2 & 82.2 \\
			\hline
			\{3, 0\} & 47.3 & 75.1 & \textbf{84.1} & 45.2 & 73.6 & 83.5 \\
			\rowcolor{blue!5}
			\{3, 1\} & \textbf{48.1} & \textbf{76.4} & \textbf{84.1} & \textbf{46.5} & \textbf{75.1} & \textbf{83.9} \\
			\{3, 3\} & 46.8 & 74.5 & 83.9 & 46.0 & 74.2 & 83.3\\
			\{3, 5\} & 46.4 & 72.8 & 83.4 & 46.4 & 74.2 & 83.6 \\
			\{3, 7\} & 46.1 & 72.9 & 83.9 & 45.1 & 72.2 & 82.9 \\
			\{3, 9\} & 45.9 & 72.4 & 83.5 & 46.3 & 73.7 & 83.4 \\
			\hline
			\{0, 1\} & 45.5 & 74.3 & 83.4 & 44.7 & 72.9 & 83.0 \\
			\{1, 1\} & 46.7 & 75.4 & \textbf{84.4} & 45.7 & 73.5 & 83.0 \\
			\rowcolor{blue!5}
			\{3, 1\} & \textbf{48.1} & \textbf{76.4} & 84.1 & \textbf{46.5} & \textbf{75.1} & \textbf{83.9} \\
			\{5, 1\} & 47.5 & 74.5 & \textbf{84.4} & 46.3 & 74.3 & 83.8 \\
			\{7, 1\} & 45.8 & 73.7 & 84.2 & 45.7 & 73.9 & 83.5 \\
			\{9, 1\} & 46.7 & 75.0 & 83.5 & 45.2 & 74.1 & 83.7 \\
			\Xhline{1.2px}
		\end{tabular}
	}
\end{table}

\textbf{Cluster Algorithm and Cluster Number $N_{p}$.} Table \ref{table: cluster methods ablation} presents the retrieval performance with different cluster algorithms in our model, including K-Means, DBSCAN, Gaussian Mixture, and DPC-KNN. Compared to other methods, DPC-KNN demonstrates better performance on most metrics. The inferior performance can be attributed to respective weaknesses. Specifically, K-Means typically fragments the temporal sequence into uniform spherical regions based on the absolute distance, undermining the semantic completeness of continuous fast-action clips. The adaptive radius optimization of DBSCAN inevitably classifies a large portion of sparse frames as outliers and discards them, thereby depriving vital long-tail clues. The multimodal sequences possess highly non-linear temporal characteristic, which structurally violates the rigid ellipsoidal Gaussian distribution assumptions of Gaussian Mixture. Therefore, DPC-KNN is more suitable for text and video tokens clustering. To achieve end-to-end training, we adopt the forward structural masking and backward attentional feature propagation mechanisms. The former leverages DPC-KNN wrapped in gradient-free scopes to adaptively generate structure masks, whereas the latter performs fully differentiable tensor aggregations via continuous linear index-addition operations. This decoupled design isolates non-differentiable graph indexing from representation learning, thereby guaranteeing robust training convergence and excellent optimization stability without performance oscillations.

Table \ref{table: cluster numbers ablation}
presents the retrieval performance for different cluster numbers $N_{p}$. For simplicity, we set $N_{p}=N_{c}$, which indicates the phrases and clips obtained from words and frames are with the same length. We find that the \textit{t2v} R@1 is firstly improved from 47.0 to 47.9 before reaching the saturation point (\textit{i.e.,} $N_{p}=6$), and then begins to decline. The main reason is that too few phrases (clips) cannot contain diverse semantic information in the texts (videos) while too many phrases (clips) may inevitably introduce redundancy or distractions, thus degrading the retrieval performance slightly. Besides, as the cluster number $N_{p}$ increases, the computational complexity (FLOPs) of the model also increases. We set $N_{p}=6$ for all datasets to achieve a balance between effectiveness and efficiency. 

\begin{table}
	\caption{\textbf{Ablation study of different cluster algorithms.}}
	\label{table: cluster methods ablation}
	\centering
	\normalsize
	\resizebox{0.5\textwidth}{!}{
		\renewcommand{\arraystretch}{1.05} 
		\begin{tabular}{l|ccc|ccc}
			\Xhline{1px}
			\multirow{2}*{Method} & \multicolumn{3}{c|}{Text-to-Video} & \multicolumn{3}{c}{Video-to-Text} \\
			\cline{2-4} \cline{5-7} 
			& R@1$\uparrow$  & R@5$\uparrow$  & R@10$\uparrow$
			& R@1$\uparrow$  & R@5$\uparrow$  & R@10$\uparrow$ \\
			\hline
			K-Means & 47.4 & 75.5 & 83.2 & 46.2 & 74.7 & \textbf{84.3} \\
			DBSCAN & 45.9 & 74.8 & 83.3 & 46.3 & 73.8 & 83.1 \\
			Gaussian & 46.8 & 72.7 & 82.8 & 46.1 & 72.5 & 82.9 \\
			\rowcolor{blue!5}
			DPC-KNN & \textbf{48.1} & \textbf{76.4} & \textbf{84.1} & \textbf{46.5} & \textbf{75.1} & 83.9 \\
			\Xhline{1px}
		\end{tabular}
	}
\end{table}

\begin{table}
	\caption{\textbf{Ablation study of cluster number $N_{p}$.}}
	\label{table: cluster numbers ablation}
	\centering
	\normalsize
	\resizebox{0.5\textwidth}{!}{
		\renewcommand{\arraystretch}{1} 
		\begin{tabular}{c|ccc|ccc|c}
			\Xhline{1px}
			\multirow{2}*{\makecell[c]{$N_{p}$}} & \multicolumn{3}{c|}{Text-to-Video} & \multicolumn{3}{c|}{Video-to-Text} & \multirow{2}*{FLOPs$\downarrow$} \\
			\cline{2-4} \cline{5-7} 
			& R@1$\uparrow$  & R@5$\uparrow$  & R@10$\uparrow$
			& R@1$\uparrow$  & R@5$\uparrow$  & R@10$\uparrow$ \\
			\hline
			2 & 47.0 & 74.8 & 83.9 & 45.3 & 72.9 & \textbf{83.9} & \textbf{36.75G}\\
			4 & 47.9 & 74.3 & \textbf{84.2} & 45.5 & 73.3 & 82.9 & 36.79G \\
			\rowcolor{blue!5}
			6 & \textbf{48.1} & \textbf{76.4} & 84.1 & 46.5 & \textbf{75.1} & \textbf{83.9} & 36.84G \\
			8 & 46.4 & 74.2 & 83.6 & 44.5 & 72.6 & 82.9 & 36.88G \\
			10 & 46.7 & 75.3 & \textbf{84.2} & \textbf{47.2} & 74.2 & 83.4 & 36.93G \\
			\Xhline{1px}
		\end{tabular}
	}
\end{table}

\textbf{Nearest Neighbors Number $K$.} In Table \ref{table: nearest neighbors ablation}, we investigate the impact of varying the nearest neighbors number $K$ in DPC-KNN. During the token clustering process, selecting a suitable $K$ is critical: a small $K$ lacks the contextual relationships modeling and produces fragmented clustered features, while a large $K$ blurs the clustered features and lacks discriminative power. To ascertain the optimal value of $K$, we set $K=\{2,3,4,5\}$. From the table, we notice that $K=3$ performs best and serves as the number of nearest neighbors in our experiments.

\begin{table}
	\caption{\textbf{Ablation study of number of nearest neighbors $K$.}}
	\label{table: nearest neighbors ablation}
	\centering
	\resizebox{0.5\textwidth}{!}{
		\renewcommand{\arraystretch}{1} 
		\begin{tabular}{l|ccc|ccc}
			\Xhline{1px}
			\multirow{2}*{$K$} & \multicolumn{3}{c|}{Text-to-Video} & \multicolumn{3}{c}{Video-to-Text} \\
			\cline{2-4} \cline{5-7} 
			& R@1$\uparrow$  & R@5$\uparrow$  & R@10$\uparrow$
			& R@1$\uparrow$  & R@5$\uparrow$  & R@10$\uparrow$ \\
			\hline
			2 & 47.7 & 75.7 & 83.9 & 42.3 & 72.1 & 83.1 \\
			\rowcolor{blue!5}
			3 & \textbf{48.1} & \textbf{76.4} & 84.1 & \textbf{46.5} & \textbf{75.1} & 83.9 \\
			4 & 46.2 & 74.8 & \textbf{84.4} & 45.4 & 74.9 & \textbf{84.5} \\
			5 & 46.3 & 75.6 & 84.1 & 45.6 & 73.4 & 84.0 \\
			\Xhline{1px}
		\end{tabular}
	}
\end{table}

\begin{figure*}[!t]
	\centerline{\includegraphics[width=\textwidth]{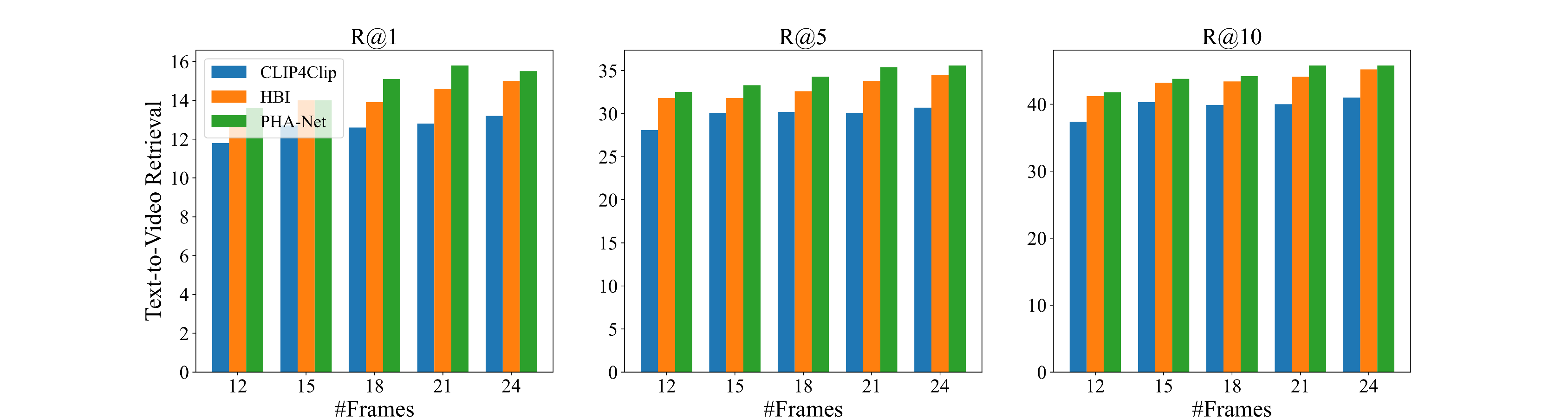}}
	\caption{\textbf{Ablation study of sampling frames $N_{f}$.}  Under different sampling frames settings, $N_{f}=\{12, 15, 18, 21, 24\}$, we compare text-to-video retrieval performance of PHA-Net with CLIP4Clip \cite{luo2022clip4clip} and HBI \cite{jin2023video} on the Charades \cite{sigurdsson2016hollywood} dataset. }
	\label{fig: bar_visualization}
\end{figure*}

\textbf{Sampling Frames $N_{f}$.} Fig. \ref{fig: bar_visualization} discusses the effect of varying frames counts $N_{f}$ on the Charades dataset. We compare our PHA-Net with two SOTA models, CLIP4Clip \cite{luo2022clip4clip} and HBI \cite{jin2023video}, and report performance with $N_{f}=\{12,15,18,21,24\}$. As observed, PHA-Net brings a notable performance enhancement under different $N_{f}$ values. Owing to the prototype-based hierarchical alignment, our approach exhibits robustness to different configurations of input videos. Since too large $N_{f}$ notably increases the resource consumption of the model, we set $N_{f}=12$ for efficient processing and fair comparisons with other methods. Notably, when handling longer videos, such as those in ActivityNet \cite{caba2015activitynet}, $N_{f}$ can be appropriately increased to achieve better performance.

\begin{figure}[!t]
	\centerline{\includegraphics[width=0.5\textwidth]{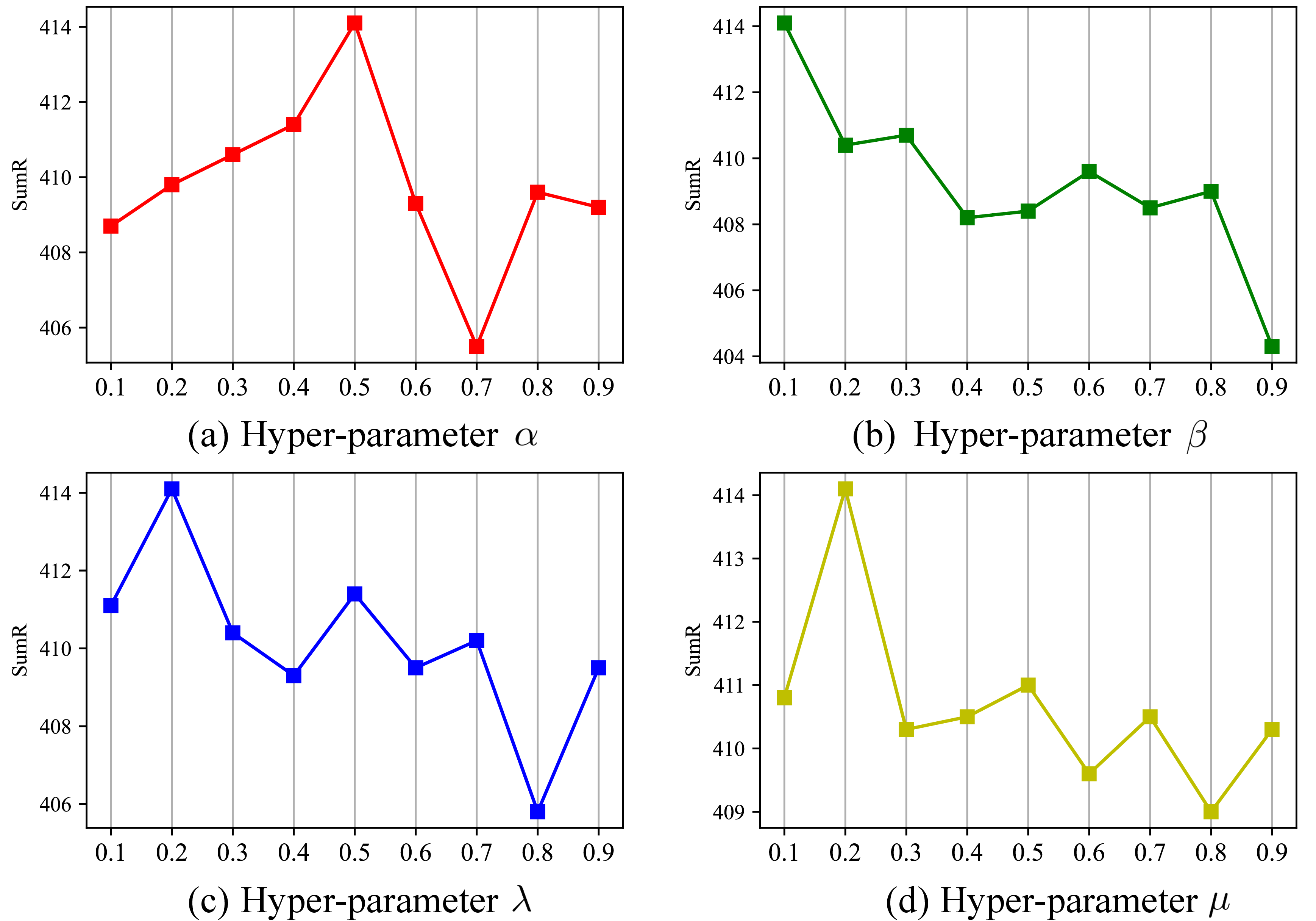}}
	\caption{\textbf{Ablation study of hyper-parameters $\alpha$ and $\beta$ in Eq. \ref{Eq: hierarchical contrastive loss} as well as $\lambda$ and $\mu$ in Eq. \ref{Eq: total loss}. }}
	\label{fig: param_ablation}
\end{figure}

\textbf{Hyper-parameters $\alpha$, $\beta$, $\lambda$, and $\mu$.} In Eq. \ref{Eq: hierarchical contrastive loss}, the hyper-parameters $\alpha$ and $\beta$ trade off $\mathcal{L}_{\text{CL}}^{I}$, $\mathcal{L}_{\text{CL}}^{L}$, and $\mathcal{L}_{\text{CL}}^{G}$. We evaluate the scale range settings $\alpha \in [0.1,0.9]$ and $\beta \in [0.1,0.9]$ as shown in Fig. \ref{fig: param_ablation} (a) and Fig. \ref{fig: param_ablation} (b), respectively. From Fig. \ref{fig: param_ablation} (a), we observe that the performance continues to grow with the increase of $\alpha$ and goes to saturation at $\alpha=0.5$ (\textit{i.e.,} 414.1 SumR). For $\beta$, the model performs best at 0.1 before decreasing with larger values. We also conduct experiments to find optimal values of hyper-parameters $\lambda$ and $\mu$ in Eq. \ref{Eq: total loss}. As shown in Fig. \ref{fig: param_ablation} (c) and Fig. \ref{fig: param_ablation} (d), the performance gets peaked at $\lambda=0.2$ and $\mu=0.2$. Therefore, we figure out the best value of $\alpha$, $\beta$, $\lambda$, $\mu$ as 0.5, 0.1, 0.2, 0.2 in practice.

\begin{table}[!t]
	\caption{\textbf{Evaluation of the computational costs with a single NVIDIA GeForce RTX 3090 GPU.} Here, the inference time is for per video evaluation. All models adopt CLIP-ViT-B/32 backbone with 64 mini-batch sizes for fair comparisons with others. The best and second best results are in bold and underlined, respectively.}
	\label{table: ablation study of efficiency analysis}
	\centering
	\normalsize
	\resizebox{0.5\textwidth}{!}{
		\renewcommand{\arraystretch}{1.1}
		\begin{tabular}{l|ccccc}
			\Xhline{1.25px}
			Method & \makecell[c]{Training\\Time$\downarrow$} & FLOPs$\downarrow$ & \makecell[c]{Inference\\Time$\downarrow$} & \makecell[c]{Inference\\Memory$\downarrow$} & SumR$\uparrow$ \\
			\hline
			HBI \cite{jin2023video} (Base) & 3h 05min & 36.38G & 541.6ms & 2971.75MB & 405.3 \\
			\hline
			CLIP4Clip \cite{luo2022clip4clip} & 3h 18min & \textbf{36.27G} & \textbf{59.3ms} & \underline{2869.89MB} & 391.7\\
			X-Pool \cite{gorti2022x} & 3h 15min & 37.34G & 346.6ms & \textbf{2837.16MB} & 403.6 \\
			X-CLIP \cite{ma2022x} & 3h 32min & \textbf{36.27G} & \underline{76.2ms} & 2942.48MB & 406.3 \\
			DRL \cite{wang2022disentangled} & 3h 10min & \underline{36.28G} & 95.0ms & 2940.17MB & 408.3 \\
			UCoFiA \cite{wang2023unified} & \underline{2h 21min} & 36.43G & 134.8ms & 3037.43MB & 409.4 \\
			TC-MGC \cite{jing2025tc} & 2h 37min & 37.39G & 673.1ms & 2947.50MB & \underline{410.1} \\ 
			\rowcolor{blue!5}
			\textbf{PHA-Net (Ours)} & \textbf{1h 56 min} & 36.84G & 407.0ms & 3044.35MB & \textbf{414.1} \\
			\Xhline{1.25px}
		\end{tabular}
	}
\end{table}

\textbf{Computational Costs Analysis.} Considering that PHA-Net needs to align cross-modal semantics at hierarchical levels, we report the detailed computational costs of our method and several recent state-of-the-art methods, including training time, FLOPs, inference time, and inference memory as shown in Table \ref{table: ablation study of efficiency analysis}. For a more rigorous comparison, we refine all metrics except training time from integer to one decimal place. Besides, to validate the effectiveness of our approach, we add a column to Table \ref{table: ablation study of efficiency analysis} about the overall SumR comparison. From the table, we elaborate on some important observations. (i) As for overall retrieval performance, PHA-Net attains the optimal SumR of 414.1 compared to other methods. (ii) In terms of the training stage, PHA-Net achieves the minimal time cost. We think the main reason is that the modality-shared learnable prototypes can jointly optimize textual and visual representations, thus stabilizing training and accelerating convergence better. (iii) In terms of the inference stage, compared to the baseline HBI, despite a slight increase in FLOPs and memory usage, our inference time is lower. The extra resource consumption arises from the increased length of individual-level tokens and added transformer blocks. However, due to fewer tokens at local and global levels, PHA-Net exhibits faster inference speed. (iv) The cross-attention-based methods like X-Pool and TC-MGC are significantly slower. When the hierarchical framework is equipped with multiple cross-attention modules, the inference speed will be severely decreased, negatively affecting real-time retrieval capability. From this point of view, our method achieves a better trade-off between superior performance and acceptable computational costs.
The above analysis demonstrates the effectiveness and efficiency of PHA-Net in practical retrieval.
 
\begin{figure*}[!t]
	\centerline{\includegraphics[width=\textwidth]{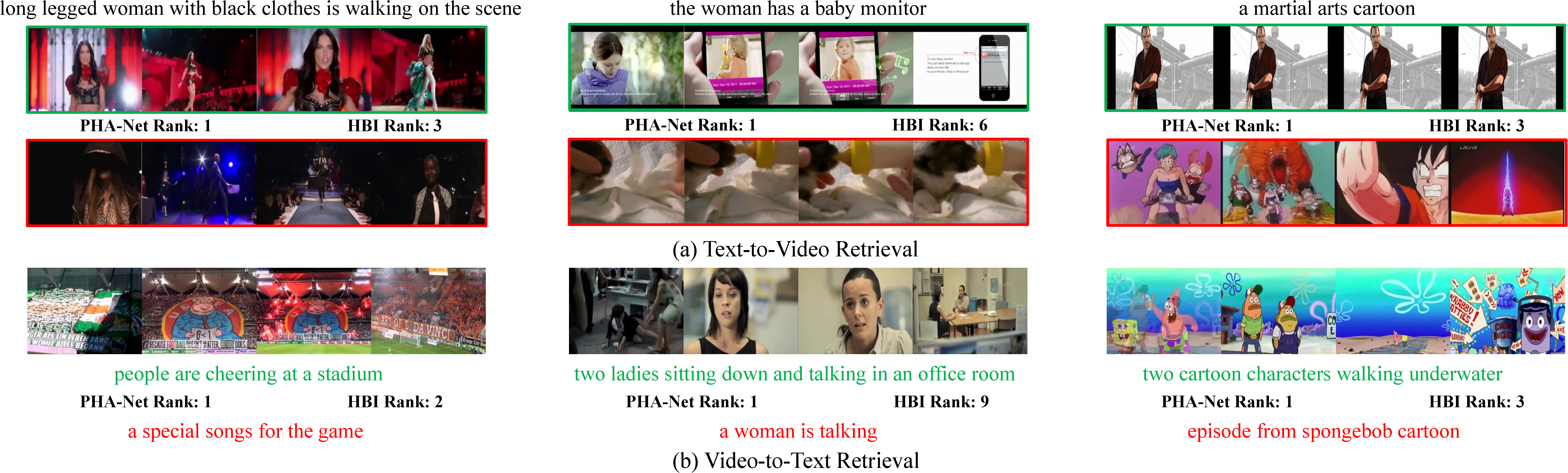}}
	\caption{\textbf{Visualization of retrieval results of our PHA-Net and HBI on the MSR-VTT dataset.} Given the text or video query, we provide the top-1 retrieved results of each method, with ground-truth and others in green and red. Note that the retrieval ranks of ground-truth in PHA-Net and HBI are shown under the ground-truth.}
	\label{fig: t2v_v2t_visualization}
\end{figure*}

\begin{figure}[!t]
	\centerline{\includegraphics[width=0.5\textwidth]{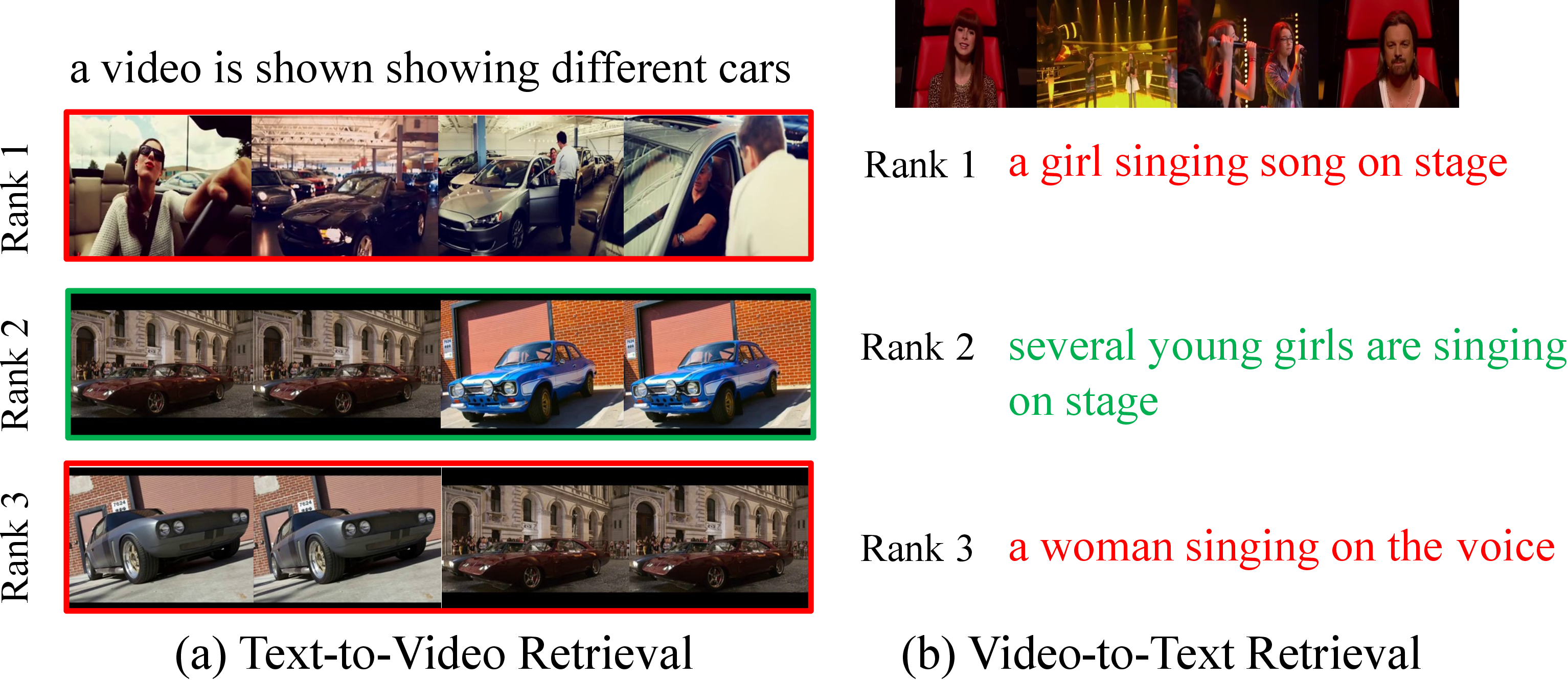}}
	\caption{\textbf{Visualization of failure analysis on the MSR-VTT dataset.} The top-1 retrieved results are not the given ground-truth, with ground-truth in green and others in red.}
	\label{fig: failure_visualization}
\end{figure}

\begin{figure*}[!t]
	\centerline{\includegraphics[width=0.9\textwidth]{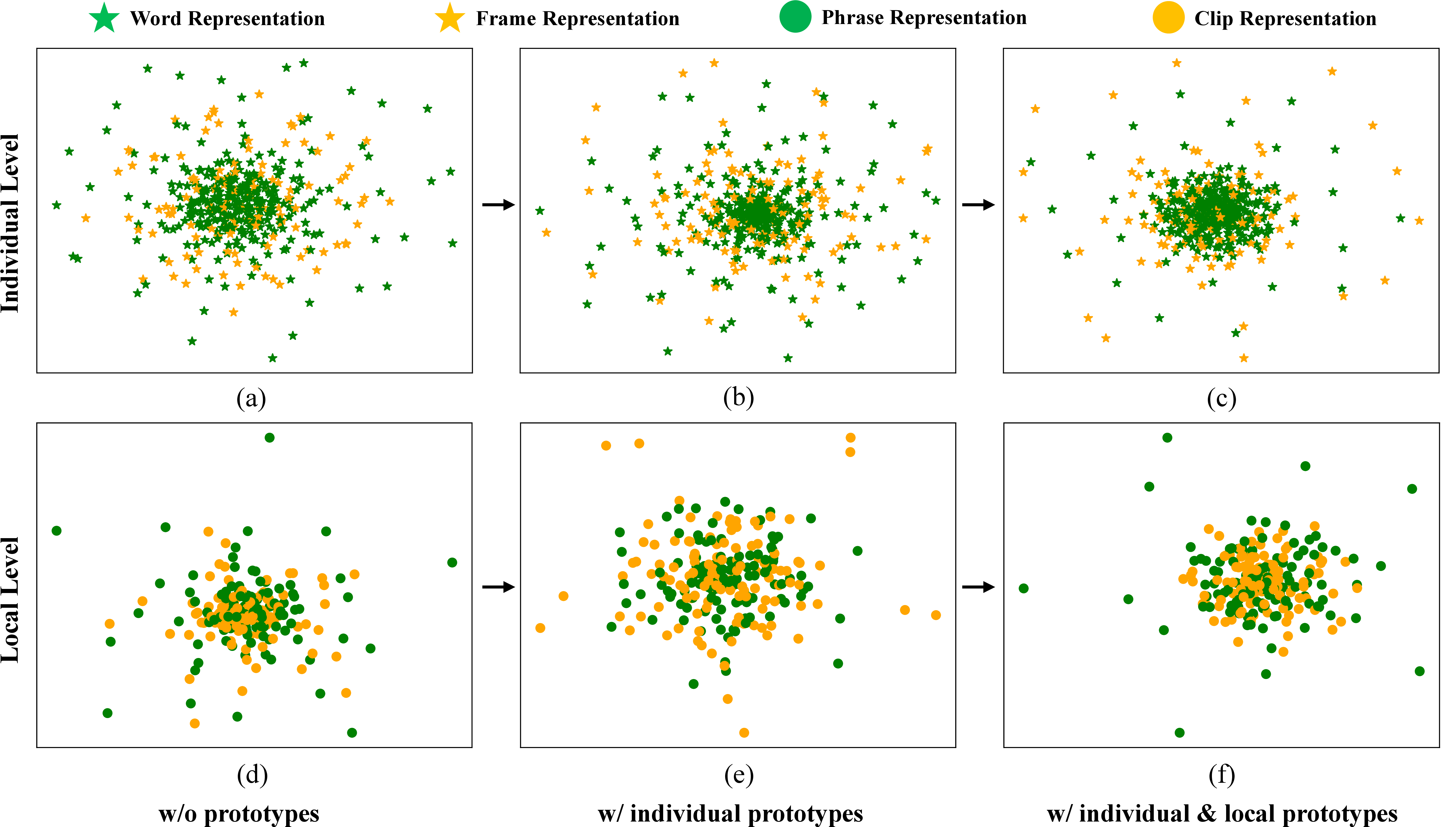}}
	\caption{\textbf{The t-SNE visualization of text and video features at individual and local levels.} Left: the results of the method without prototypes. Middle: the results of the method with individual prototypes. Right: the results of the method with individual and local prototypes. }
	\label{fig: tsne_visualization}
\end{figure*}

\begin{figure*}[!t]
	\centerline{\includegraphics[width=0.9\textwidth]{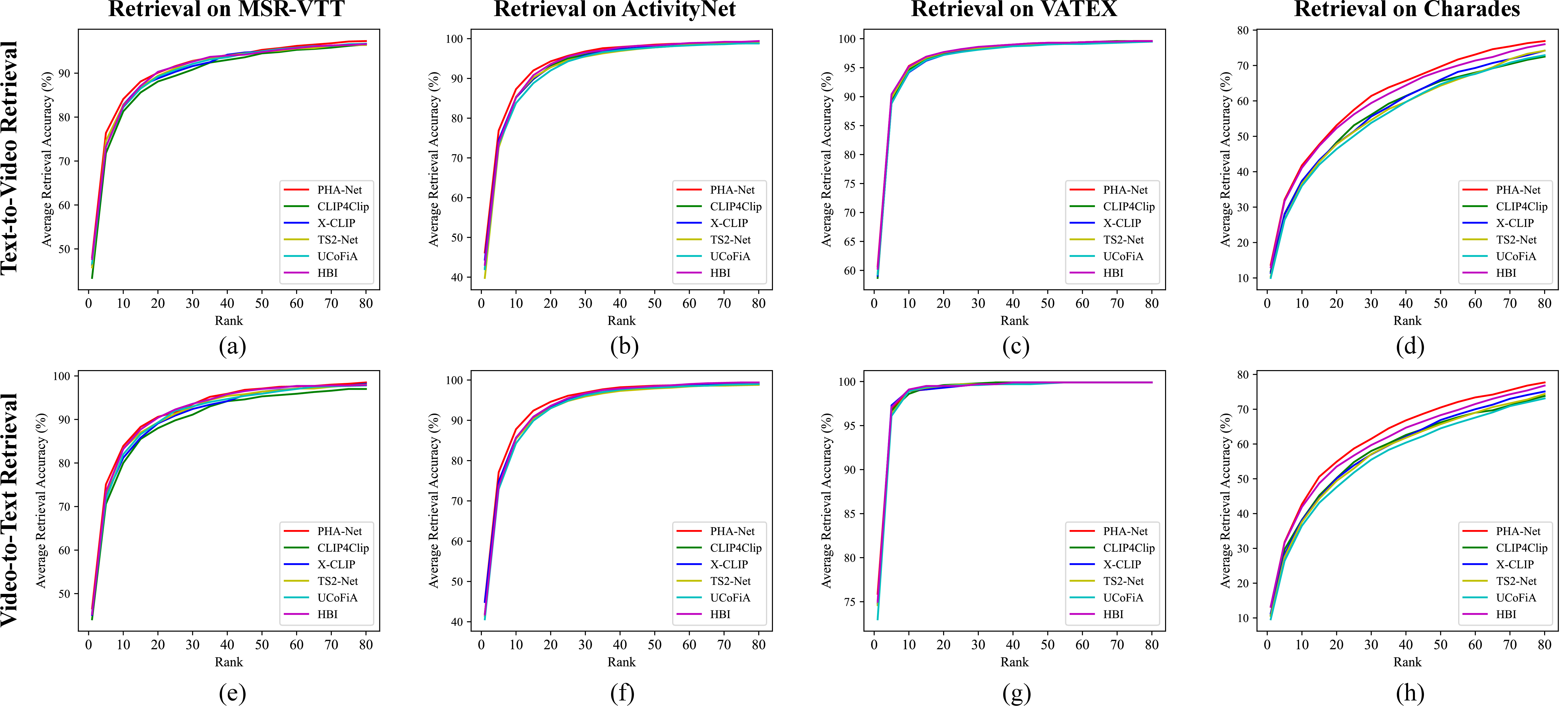}}
	\caption{\textbf{CMC curves comparison between our method and other methods for both the text-to-video retrieval and video-to-text retrieval tasks.} The curves from left to right correspond to MSR-VTT, ActivityNet, VATEX, and Charades, respectively.  }
	\label{fig: cmc_curve}
\end{figure*}

\subsection{Qualitative Analysis}
\textbf{Retrieval Results.} To qualitatively illustrate the effectiveness of the proposed approach, we show some retrieval results of our PHA-Net and HBI from the MSR-VTT dataset as shown in Fig. \ref{fig: t2v_v2t_visualization}. From the text-to-video retrieval results in the top part, the first example reflects that our PHA-Net can recognize visual entity with composite textual clues, successfully locating ``long legged woman'' in frames 2 and 4. Likewise, the second example shows our model's capability of understanding complex semantic concept of ``baby monitor'', which is quite important for retrieving correct videos. This is because the varying prototypes can effectively identify different semantics hidden in the compound text, thus achieving accurate text-video matching through prototype-level alignment. In the third example, only the video retrieved by our model fits the abstract semantics ``martial arts'', showing the detailed semantic perception of learnable prototypes. From the video-to-text retrieval results in the bottom part, in the first and third examples, we find that the retrieved results of HBI are all overall descriptions about video queries but only PHA-Net captures static entities like ``people'', ``stadium'', ``cartoon characters'' as well as actions like 
``cheering'' and ``walking''. In the second example, although HBI understands partial action of ``talking'', it fails to capture other action, such as ``sitting down'' in the frame 4. Both text-to-video and video-to-text samples demonstrate the superiority of modality-shared prototypes for correct video and text retrieval.

We also visualize the failure cases in Fig. \ref{fig: failure_visualization}, where the ground-truth does not rank first in the retrieved list. In the left example, PHA-Net retrieves the video with cars, but they are not different, which is ascribed to the aforementioned PCA in the token merge module. Specifically, the sharp scene transitions in the ground-truth video are smoothed by the 1D-Conv layer, resulting in the semantic dilution of high-frequency temporal clues after DPC-KNN clustering. When these merged tokens are treated as $K$ and $V$, the output tokens become less discriminative, thereby yielding a sub-optimal similarity score relative to the query text. In the right example, PHA-Net retrieves the text that describes the singing action, ignoring the critical semantics of the number of girls. The possible reason is that the limited prototypes are insufficient to perceive the micro-level quantity attribute, thereby impairing the fine-grained discriminative power of individual-level alignment. In summary, these failure cases reveal the limitations of our approach and provide valuable insights for future research.

\textbf{t-SNE Visualization.} Fig. \ref{fig: tsne_visualization} shows the distribution of text and video features at individual as well as local levels using the t-SNE \cite{van2008visualizing} visualization tool. Notably, the learnable prototypes are not taken into account when visualizing the feature distribution. From Fig. \ref{fig: tsne_visualization} (b), we observe that individual prototypes can effectively narrow the semantic gap across modalities and facilitate cross-modal alignment. We also find that the introduction of local prototypes can align words and frames better as illustrated in Fig. \ref{fig: tsne_visualization} (c), which may be attributed to the gradient backpropagation from local level to individual level. For the local-level features in the bottom part, as shown in Fig. \ref{fig: tsne_visualization} (f), our approach gathers phrases and clips together better with the local prototypes' guidance, further demonstrating the effectiveness of modality-shared prototypes in reducing the modality gap. 

\textbf{CMC Curves Evaluation.} Fig. \ref{fig: cmc_curve} shows the Cumulative Match Characteristic (CMC) curves of model evaluation in all four datasets. It can be observed that our PHA-Net surpasses all compared models on four datasets, which clearly demonstrates the effectiveness of our approach. Specifically, our PHA-Net consistently improves the performance by a large margin on the MSR-VTT, ActivityNet, and Charades datasets. As for the VATEX dataset, our PHA-Net shows relatively minor performance gains, which may be attributed to the diversity of video descriptions. We believe more learnable prototypes are needed to handle such diverse scenarios and achieve significant performance improvements.

\begin{figure*}[!t]
	\centerline{\includegraphics[width=\textwidth]{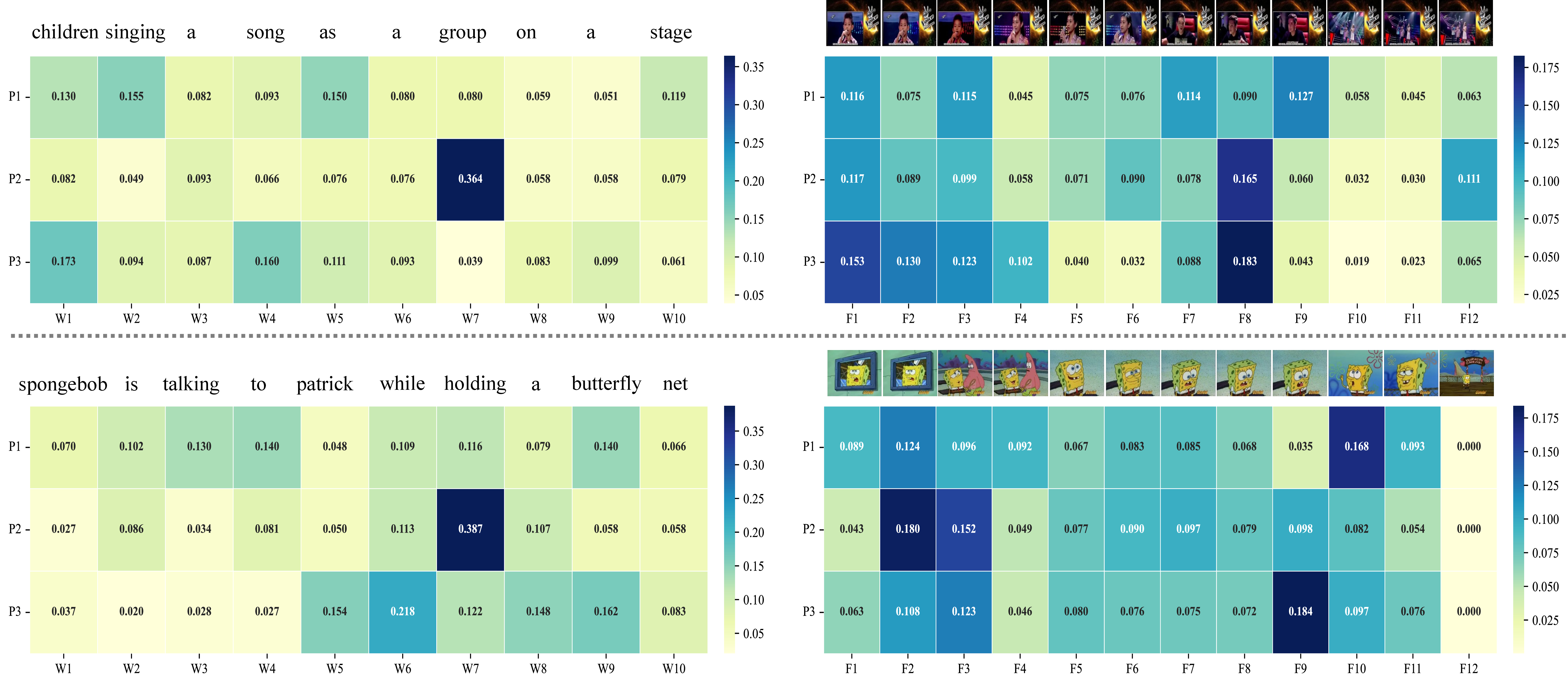}}
	\caption{\textbf{Heatmap of attention weights between prototype and words as well as prototypes and frames.} We take video9303 and video8446 in the MSR-VTT dataset as examples. Darker colors indicate higher similarity, while lighter colors indicate lower similarity. }
	\label{fig: proto_correlation}
\end{figure*}

\textbf{Prototype Correlation.} The core idea of our method is to use modality-shared prototypes for capturing shared semantics. To further analyze and understand this capability, we provide the heatmap visualization of prototype-word and prototype-frame attention weights in Fig. \ref{fig: proto_correlation}. Notably, the attention weights are derived from the second transformer layer in the transformer blocks. As observed, the words and frames with 
similar semantics are tightly connected with the same prototype, and different prototypes focus on diverse semantic regions. In the top example, the words ``as'' and ``singing'' are most associated with the first prototype, and the frames 1, 3, 7, and 9 are uniformly assigned to this prototype. This reveals the effectiveness of first prototype in representing global contextual information. The second prototype exhibits the highest relevance with the word ``group'', and the frames 8 and 12 categorized under this prototype also contain analogous semantic content, showing the successful perception of plurality semantics. Furthermore, the words ``children'' and ``song'' are assigned to the third prototype, and frames belonging to this prototype also represent the ``song-related'' visual scenarios, demonstrating the advantage of capturing action semantics. In the bottom example, three prototypes are used as global context, motion relation and action entity indicators, respectively. These examples fully illustrate that our proposed method can effectively bridge the modality gap via the shared semantics. 

\begin{figure}[!t]
	\centerline{\includegraphics[width=0.5\textwidth]{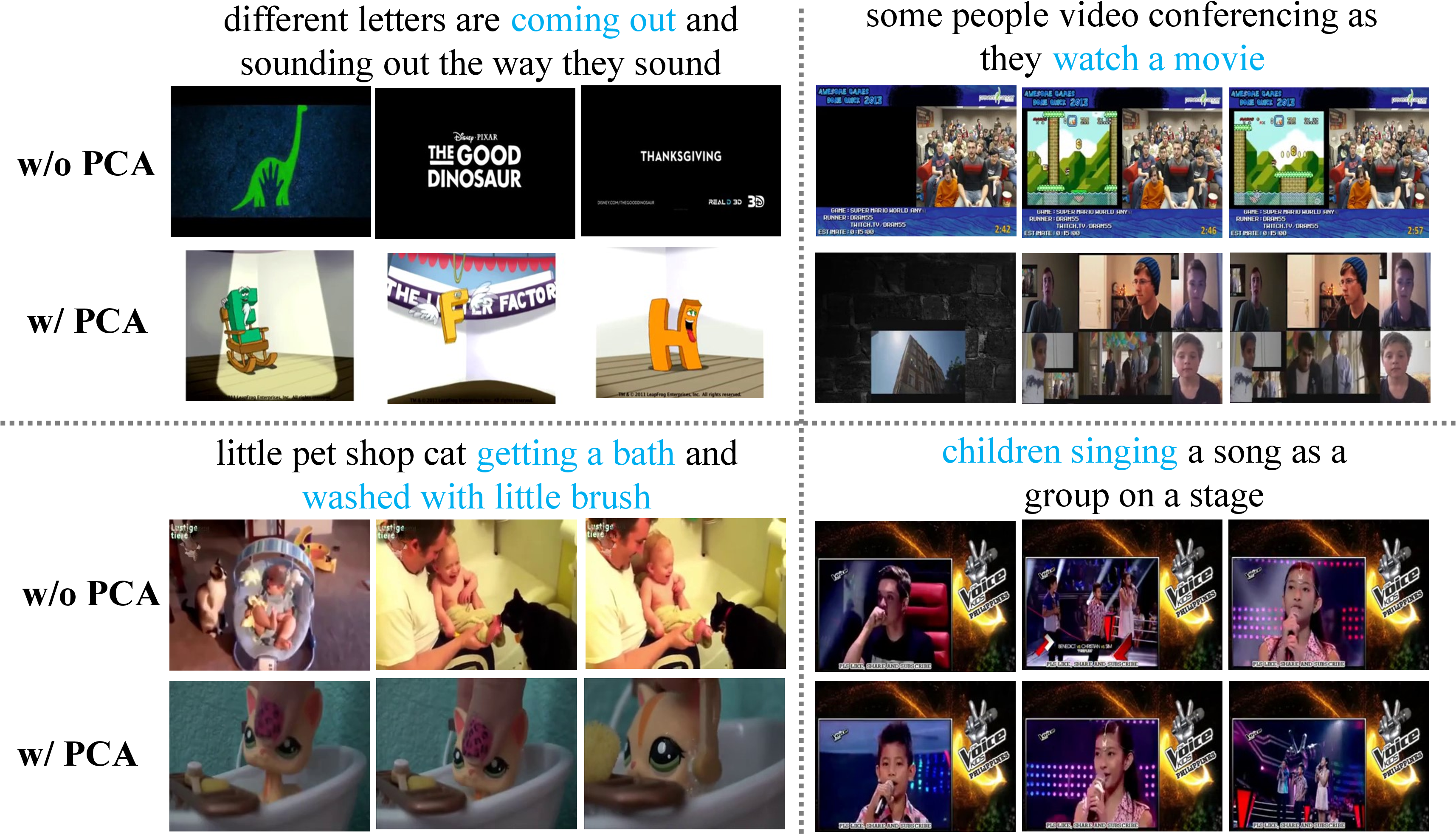}}
	\caption{\textbf{Visualization of the top-1 text-to-video retrieval results with and without the prototype-supported cross-attention (PCA) on the MSR-VTT dataset, with related words in blue.} }
	\label{fig: w_wo_pca_t2v_visualization}
\end{figure}

\begin{figure}[!t]
	\centerline{\includegraphics[width=0.5\textwidth]{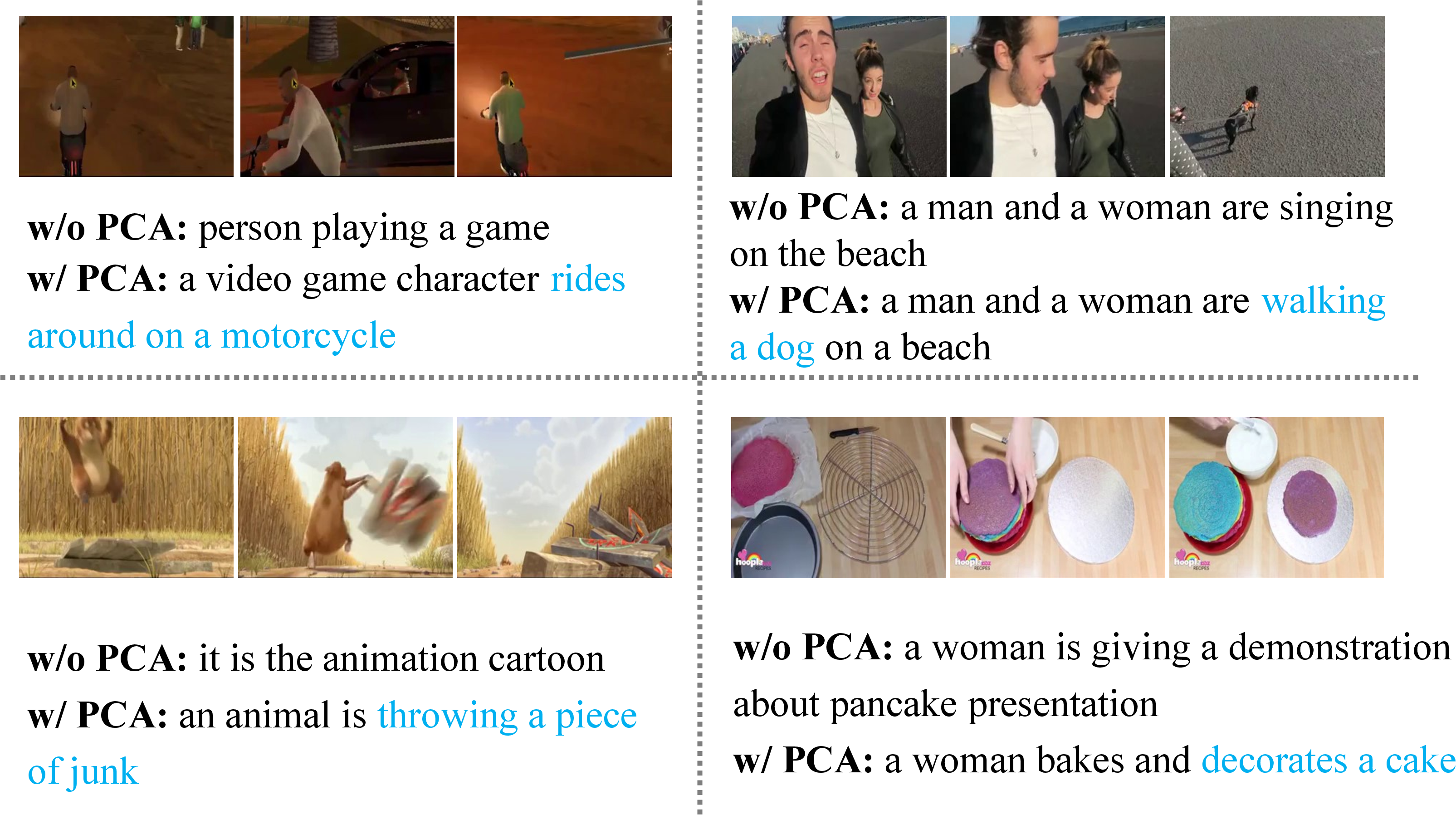}}
	\caption{\textbf{Visualization of the top-1 video-to-text retrieval results with and without the prototype-supported cross-attention (PCA) on the MSR-VTT dataset, with related words in blue.} }
	\label{fig: w_wo_pca_v2t_visualization}
\end{figure}

\textbf{Prototype-supported Cross-attention.} The comparison of text-to-video and video-to-text retrieval results without and with the proposed prototype-supported cross-attention (PCA) is illustrated in Fig. \ref{fig: w_wo_pca_t2v_visualization} and Fig. \ref{fig: w_wo_pca_v2t_visualization}. For better observation,  we uniformly sample 3 frames per video. As can be seen, integrating PCA into the token merge module clearly improves the model's matching accuracy by capturing specific entities (\textit{e.g.,} ``little brush'' and ``a dog'') and perceiving complex events (\textit{e.g.,} ``watch a movie'' and ``decorates a cake'') in text and video modalities. The experimental findings strongly prove the effectiveness of PCA in our proposed method.

\section{Conclusion}
\label{Sec: Conclusion}
In this paper, we analyze the inherent semantic mismatch problem in existing hierarchical text-video retrieval and propose a novel prototype-based hierarchical alignment network (PHA-Net) to align text and video representations for text-video retrieval. We introduce multiple modality-shared trainable prototypes at individual level and local level for joint optimization of textual and visual representations, thus achieving comprehensive hierarchical cross-modal alignment. In the token merge module, to better utilize the imbalanced semantic distribution among clustered tokens, we integrate the prototype semantic guidance into the merging process for the enhancement of tokens with strong  semantics and suppression of those with weak semantics. Besides, we design an auxiliary prototype contrastive loss to constrain the textual prototype close to  its corresponding visual prototype and far from others, guaranteeing advantage in prototype diversity. Extensive results on four benchmark datasets demonstrate the effectiveness and efficiency of our approach. In the future, it would be interesting to generate flexible prototype number according to various complexity of video content as the static prototype number leads to poor cross-scenario generalization capability. The classic dynamic prototype network may be a feasible solution. Additionally, we leave our method extension to other multi-modal applications like multi-modality medical image analysis and RGB-D saliency detection, as part of future work.

\section*{Acknowledgments}
This research was partially funded by the China National Key  R\&D Research Program (2020YFB1711200) and (2019YFB1705801).



\end{document}